\documentclass{aa}

\usepackage{natbib}
\usepackage{lscape}
\usepackage{graphicx}
\usepackage{txfonts}
\usepackage{adjustbox}
\usepackage{threeparttable}
\usepackage{lscape}
\usepackage{color}
\usepackage[dvipsnames]{xcolor}
\usepackage{longtable}
\usepackage{ulem}
   \usepackage[	colorlinks=true, 
   				citecolor=blue,
   				urlcolor=blue,
   				linkcolor=blue,
               pdftex,   % for pdflatex
               hyperindex=true,   % Make the text of index entries into hyperlinks
               bookmarks=true,   % Create bookmarks for the contents
               bookmarksopen=true,   % Expand the bookmark tree
               bookmarksnumbered=true]   % Include section numbers
               {hyperref}   % To turn references, figures, titles... into hyperlinks. 'hyperref' has to be the LAST package, except if one uses the package 'glossaries'
\defcitealias{2018A&A...610A..29K}{I}
\defcitealias{2019A&A...632A..28K}{II}

\begin{document}

   \title{Tomography of cool giant and supergiant star atmospheres}

   \subtitle{III.  Validation of the method on VLTI/AMBER observations of the Mira star S Ori\thanks{Based on ESO observing program 084.D-0595} }

   \author{Kateryna Kravchenko
          \inst{1,2}, Markus Wittkowski \inst{2}, Alain Jorissen\inst{3},
          Andrea Chiavassa \inst{4}, Sophie Van Eck \inst{3}, Richard~I.~Anderson \inst{2}, Bernd Freytag \inst{5},
           Ulli K{\"a}ufl \inst{2}}

   \institute{European Southern Observatory, Alonso de Cordova 3107, Vitacura, Casilla, 19001, Santiago de Chile, Chile   \\
   \email{kateryna.kravchenko@eso.org}
   \and
   European Southern Observatory, Karl-Schwarzschild-Str. 2, 85748 Garching bei M{\"u}nchen, Germany
   \and
   Institut d'Astronomie et d'Astrophysique, Universit\'e Libre de Bruxelles,
              CP. 226, Boulevard du Triomphe, 1050 Bruxelles, Belgium
                          \and
             Universit\'e C\^ote d'Azur, Observatoire de la C\^ote d'Azur, CNRS, Lagrange, CS 34229, 06304 Nice Cedex 4, France 
          \and
     Department of Physics and Astronomy at Uppsala University, Regementsv{\"a}gen 1, Box 516, SE-75120 Uppsala, Sweden     
            }

% \abstract{}{}{}{}{} 
% 5 {} token are mandatory
 
  \abstract
{Asymptotic giant branch (AGB) stars are characterized by substantial mass loss, whose mechanism is not fully understood yet. The knowledge of the structure and dynamics of AGB-star atmospheres is crucial to better understand the mass loss. The recently established tomographic method, that relies on the design of spectral masks containing lines forming in given ranges of optical depths in the stellar atmosphere, is an ideal technique for this purpose.}
{We aim at validating the capability of the tomographic method to probe different geometrical depths in the stellar atmosphere and at recovering the relation between optical- and geometrical-depth scales.}
{The tomographic method is applied to high-resolution spectro-interferometric VLTI/AMBER observations of the Mira-type AGB star S Ori. First, the interferometric visibilities are extracted at wavelengths contributing to the tomographic masks and fitted to those computed from a uniform disk model. This allows the measurement of the geometrical extent of the atmospheric layer probed by the corresponding mask. Then, we compare the observed atmospheric extension with those measured from available 1D pulsation CODEX models and 3D radiative-hydrodynamics CO5BOLD simulations.}
{While the average optical depths probed by the tomographic masks in S~Ori decrease (with $<\log \tau_0>$~$= -0.45$, $-1.45$, and $-2.45$ from the innermost to the central and outermost layers), the angular diameters of these layers increase, from 10.59 $\pm$ 0.09 mas through 11.84 $\pm$ 0.17 mas, up to 14.08 $\pm$ 0.15 mas. A similar behavior is observed when the tomographic method is applied to 1D and 3D dynamical models.}
   {This study derives, for the first time, a quantitative relation between optical- and geometrical-depth scales when applied to the Mira star S~Ori, or to 1D and 3D dynamical models. In the context of Mira-type stars, the knowledge of the link between the optical and geometrical depths opens the way to derive the shock-wave propagation velocity, which can not be directly observed in these stars.} 

\titlerunning{Tomography of cool giant and supergiant star atmospheres}   
\authorrunning{Kravchenko et al.}  
\maketitle

%
%-------------------------------------------------------------------

\section{Introduction}

Asymptotic giant branch (AGB) stars represent a late stage in the evolution of low- and intermediate-mass stars (i.e., with a mass lower than 8~M$_{\odot}$, to avoid carbon ignition in the stellar core). They are characterized by large changes in brightness and a substantial mass loss. Mass loss from oxygen-rich (C/O~$<1$) AGB stars is driven by a complex two-step process: pulsations and convection generate large-scale non-spherical shock waves in the atmosphere which lift matter in layers that are cool enough for dust grains to form \citep{2019A&A...623A.158H}. There, radiation pressure on dust pushes it further away, and provided that the coupling between gas and dust is efficient enough (i.e., the density should be larger than some threshold), mass is driven towards infinity \citep{2018A&ARv..26....1H}. Other physical mechanisms, such as magnetic fields and Alfv{\'e}n waves \citep{2010ApJ...723.1210A}, may contribute as well. The resulting mass loss, in turn, enriches the interstellar medium with gas and dust species. Thus, convection, pulsation and dust formation are crucial to the mass-loss process. Current 1D and 3D stellar-atmosphere models are able to simulate some of these processes and provide reasonable agreement with the observed spectral features of AGB stars \citep[e.g.,][]{2016A&A...587A..12W}. However, a much stronger test of model atmospheres (with a large number of physical and numerical parameters) is  their confrontation with detailed multi-dimensional time-dependent observations. The recently established tomographic method provides a further step in this direction.

\citet{2001A&A...379..288A} developed the tomographic technique to follow the propagation of the shock wave through the atmospheres of the Mira-type AGB stars. The method is based on the construction of spectral templates (henceforth "masks"), which contain lines forming at different atmospheric depths. The cross-correlation of masks with stellar spectra allows the reconstruction of the disk-averaged velocity fields at different optical depths in the stellar atmosphere \citep[][hereafter Papers~\citetalias{2018A&A...610A..29K} and \citetalias{2019A&A...632A..28K}]{2018A&A...610A..29K,2019A&A...632A..28K}. 

So far, the tomographic method has only been used with optical depths as proxy for geometrical depths. Although current model atmospheres provide a relation between optical and geometrical depths \citep[e.g.,][]{2016A&A...587A..12W}, it may be sensitive to the stellar mass which is ill-defined since stars are undergoing a substantial mass loss. In addition, current stellar-atmosphere models experience several limitations. For example, in 1D hydrostatic model atmospheres, the dynamical processes like convection and pulsations are parameterized via macro- and micro-turbulence \citep{2008A&A...486..951G}. Current 1D and 3D dynamical models \citep{2008MNRAS.391.1994I,2011MNRAS.418..114I,2017A&A...600A.137F} are able to simulate convection and pulsation processes, but they do not yet include wind, radiation pressure and magnetic field, which may affect the structure and dynamics of the atmosphere. Thus, the knowledge of the relation between optical- and geometrical-depth scales would open a way to test and constrain the state-of-the-art hydrostatic and dynamical model atmospheres. In the context of Mira-type AGB stars, such a relation would open the way to derive unknown properties of the shock wave, like its propagation velocity. 

\citet{2016A&A...589A.130L} used state-of-the-art 1D dynamical DARWIN models and showed that the efficiency of dust formation and mass loss rate strongly depend on the pulsation phase in which the material behind the shock wave arrives in layers sufficiently cool to allow dust to form. In this context, the shock wave velocity is crucial for the arrival time of the material in the dust-formation layers. However, in current dynamical models the shock wave velocity is provided as an output and, thus, can not be adjusted. Therefore, derivation of the shock wave velocity in real stars will be an important test of dynamical model atmospheres.

Previous interferometric observations of Mira stars in the near-IR $K$ band with the ESO Very Large Telescope Interferometer (VLTI) and the AMBER instrument \citep{2007A&A...464....1P} using low- ($R \sim 35$), medium- ($R \sim 1500$) and high-spectral-resolution ($R \sim 12\,000$) modes revealed the wavelength dependence of visibility as well as diameter variations \citep{2008A&A...479L..21W,2011A&A...532L...7W,2016A&A...587A..12W,2016A&A...589A..91O}. The squared visibility amplitudes generally show maxima around 2.25~$\mu$m corresponding to the continuum layers. The (most opaque) molecular bands are linked to the minima of the squared visibility amplitudes around 2.00~$\mu$m and 2.29~$\mu$m. These minima were interpreted as due to the presence of cool and extended molecular layers \citep[at $\sim$~1.5~$R_*$ in $\rm H_2O$ and at $\sim$ 3 $R_*$ in CO;][]{2016A&A...589A..91O} in Mira-star atmospheres. This was, in turn, confirmed by 1D and 3D dynamical atmosphere models \citep{2016A&A...587A..12W}.

%--------------------------------------------------- One column table
\begin{table*}
\begin{center}
\caption[]{Summary of VLTI/AMBER observations of S~Ori retained in the final dataset.}
\label{tab:data_overview} 
%$$ 
\begin{tabular}{c c c c c c c c}
\hline \hline
\noalign{\smallskip}
Date & $\rm \Phi_{Vis}$ & Central wavelength & Bandwidth   & Projected baseline [m]  & Position angle & Seeing & Airmass    \\
  &   & [$\mu \rm m$] & [$\mu \rm m$]      &        H0-G0/G0-E0/E0-H0         &    [$^\circ$]  & [$"$]  &             \\
\noalign{\smallskip}
\hline
\noalign{\smallskip}
28-12-2009 & 0.62 & 2.288 & 0.046 &  28.05 / 14.03 / 42.09 &  253 & 0.8 & 1.30 \\     
28-12-2009 & 0.62 & 2.326 & 0.045 &  30.42 / 15.22 / 45.65 &  253 & 0.7 & 1.16 \\ 
28-12-2009 & 0.62 & 2.211 & 0.048 &  31.13 / 15.58 / 46.71 &  253 & 0.7 & 1.12 \\ 
17-01-2010 & 0.67 & 2.133 & 0.046 &  31.34 / 15.68 / 47.01 &  251 & 0.6 & 1.07 \\ 
\noalign{\smallskip}
\hline
\end{tabular}
\end{center}
\end{table*}

Our target star S Ori is a Mira variable with a pulsation period of about 414~d \citep{2017ARep...61...80S}. However, the period of S~Ori varies from cycle to cycle between about 400 and 450~d \citep{2005AJ....130..776T}. Its visual magnitude varies between 7.2 and 14.0, while the spectral type varies between M6.5e and M9.5e. S~Ori has atmospheric parameters  $T_{\rm eff} = 2950$~K \citep{2011A&A...531A..88U} and $\log g = -0.8^{+0.3}_{-0.2}$ \citep[computed from $R = 420 \pm 130$~R$_{\odot}$ derived by][and $M = 1$~M$_{\odot}$]{2008A&A...479L..21W}. The Gaia DR2 parallax of 1.86 $\pm$ 0.34 mas \citep{2016A&A...595A...1G,2018A&A...616A...1G} results in a distance of $538^{+120}_{-83}$~pc. The near-infrared $K$-band uniform disk (UD) angular diameter of S Ori has been previously measured by \citet{1996AJ....112.2147V}, \citet{2005ApJ...620..961M}, and \citet{2005ApJ...618..953B}. It ranges from 9.6 to 10.5~mas depending on the variability phase, with typical errors between 0.1 and 0.6~mas. Later on, \citet{2008A&A...479L..21W} measured UD diameters for S~Ori at phase 0.78 for different channels across the $J$, $H$ and $K$ bands using VLTI/AMBER data in low-resolution mode. The derived UD diameters of S~Ori strongly depended on wavelength, with a minimum value of 8.1~mas in the near-continuum bands and about 11~mas in water vapor and CO bands.  

In this paper, we present the first high-spectral-resolution (R~$\sim$~12~000) interferometric VLTI/AMBER observations of S~Ori. We aim at deriving, for the first time, the relation between optical- and geometrical-depth scales for S~Ori by applying the tomographic method outlined by \citet{2001A&A...379..288A} and Paper~\citetalias{2018A&A...610A..29K}. This will, in turn, validate the ability of the tomographic technique to correctly probe distinct layers in stellar atmospheres. 

The paper is organized as follows. Section~\ref{Sect:observations} describes our AMBER observations and the data-reduction process. The tomographic method is presented in Sect.~\ref{Sect:tomography} and is applied to the AMBER data, and 1D and 3D dynamical stellar-atmosphere models in Sect.~\ref{Sect:UD_fit}. The implications of our results and our conclusions are drawn in Sect.~\ref{Sect:conclusions}.

\section{Observations and data reduction}
\label{Sect:observations}

\begin{figure}
   \centering
   \includegraphics[width=8.9cm]{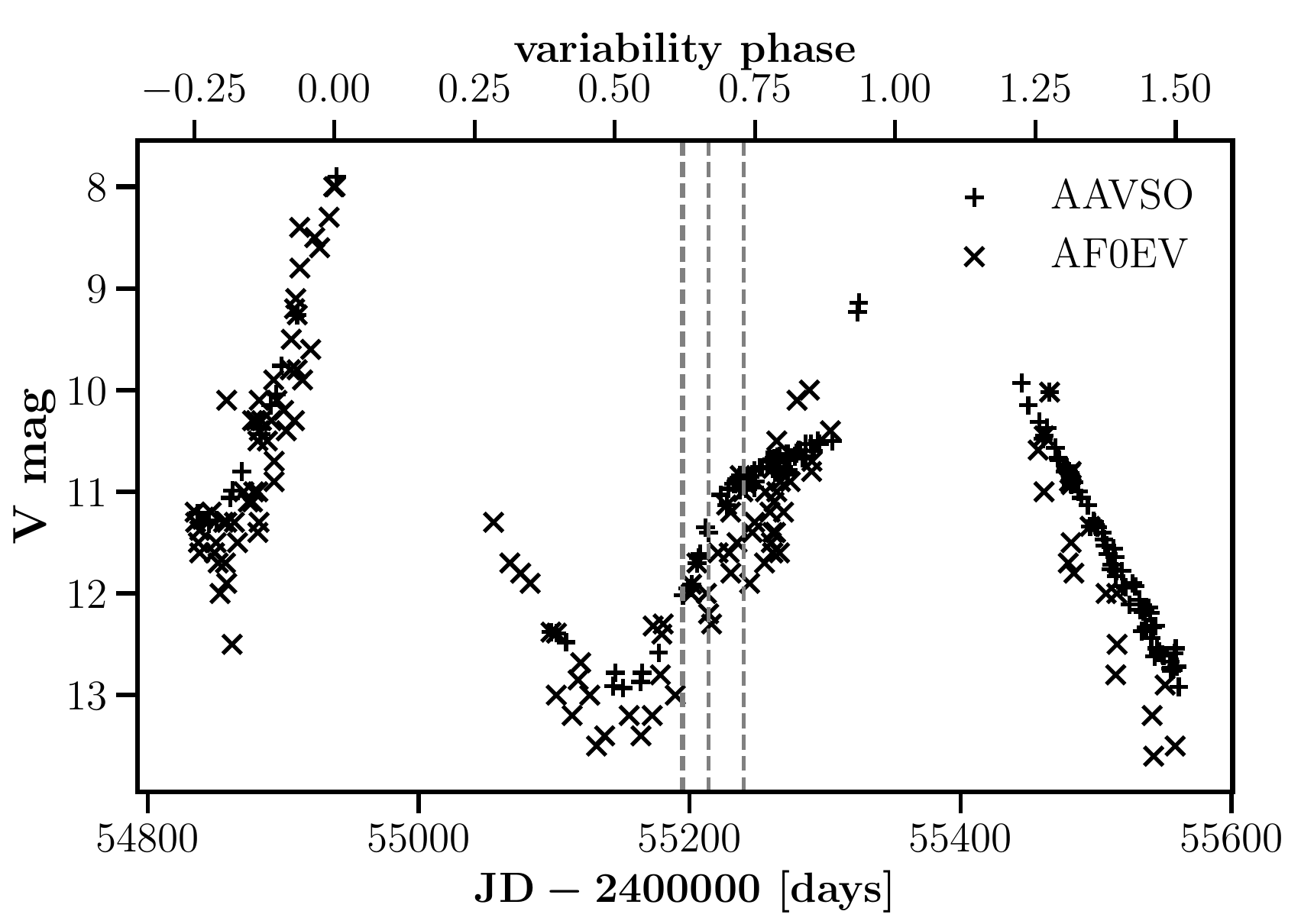}
      \caption{ Visual light curve of S Ori as a function of Julian Date (JD) and stellar variability phase. Data are taken from the AAVSO \citep{AAVSO} and AFOEV (through CDS) databases. Vertical lines correspond to the epochs of the AMBER observations.
      }
         \label{Fig:light_curve}
   \end{figure}

We observed S~Ori using the near-IR VLTI/AMBER beam combiner in high-spectral-resolution (R~$\sim$~12~000) mode. The observations were performed between December 28, 2009 and March 24, 2010 in six wavelength settings inside the $K$-band (with central wavelengths from 2.133 to 2.326~$\mu \rm m$) using the fringe tracker FINITO \citep{2008SPIE.7013E..18L}. Figure~\ref{Fig:light_curve} shows the visual light curve of S~Ori as obtained from the AAVSO\footnote{\url{https://aavso.org/}} and AFOEV\footnote{\url{http://cdsarc.u-strasbg.fr/afoev/}} databases around the dates of our observations. 

Following \citet{2015A&A...575A..50A}, the selection of the final dataset was based on the performance of the FINITO fringe tracker. During the integration of AMBER frames, random jitter may lead to a decrease of the fringe contrast (the so-called "visibility"). The factor $\rm{e}^{-\sigma^2_{\phi}}$  by which the visibility decreases is called "FINITO factor" (where $\sigma_{\phi}$ is the fringe-phase standard deviation over the frame acquisition time). In order to correct for that effect, we keep only AMBER observations characterized by similar (i.e., within 10\%) FINITO factors for the science and calibrator targets. Observations corresponding to the phase 0.73 were found to have poor FINITO performance and were thus removed from the final dataset. Table~\ref{tab:data_overview} summarizes the properties of our final dataset together with the corresponding telescope configurations, position angles, wavelength settings, optical seeing, and airmasses. For each wavelength setting, the calibrator star 31 Ori was observed shortly after or before the science target. The spectral type of the calibrator star is K4III, and its angular $K$-band UD diameter is 3.83 $\pm$ 0.40 mas \citep{2016A&A...589A.112C}.

%We performed the data reduction 
Each wavelength setting of AMBER from Table~\ref{tab:data_overview} was reduced with the latest version of the \textit{amdlib} package \citep[][release 3.0.9]{2007A&A...464...29T,2009A&A...502..705C}. First, we removed bad pixels defined in the bad pixel map (BPM) and applied the flat-field correction. Then, we averaged individual frames taken consecutively for the same source and the same wavelength setting into OI-FITS files. The selection of frames was based on three criteria: baseline flux signal-to-noise ratio (S/N), piston value (fluctuation of an optical path difference) and fringe S/N. Frames with a baseline flux S/N lower than 100 and a piston larger than 4~$\mu \rm m$ were discarded. Finally, we averaged the remaining frames keeping those with the 50\%  best fringe S/N. At the same time, 20\% and 80\% of the best frames were averaged, and the difference between these two averaged frames defines the systematic error associated with the selection of frames.      

In order to obtain correctly calibrated visibilities, an accurate wavelength calibration (not provided by {\it amdlib}) is needed both for the science and calibrator targets. For this purpose, we first performed the wavelength calibration for the (flux) spectra and then applied it to visibilities. The wavelength calibration using the telluric spectrum alone was not feasible. At the spectral resolution of 12~000, it is difficult to disentangle very complex spectral features of the science and calibrator targets and a few telluric lines present in their spectra. Using only the telluric spectrum would lead to severe ambiguities in the wavelength calibration. Thus, an absolute wavelength calibration was done using a reference spectrum that includes a high-resolution (R~$\sim$~200~000) 1D MARCS\footnote{\url{http://marcs.astro.uu.se}} \citep{2008A&A...486..951G} spectrum matching S~Ori or 31~Ori, a telluric spectrum produced with the ESO \textit{SkyCalc} tool \citep{2012A&A...543A..92N,2013A&A...560A..91J}, and the AMBER transmission curve, all convolved to the spectral resolution of AMBER \citep[see also][]{2011A&A...532L...7W}. According to the {\it amdlib} user manual, instrumental errors linked to the dispersive elements of AMBER lead to a wavelength displacement of $\sim$~10 pixels. Moreover, \citet{2010A&A...511A..51C} showed that for the low-resolution AMBER observations a constant wavelength stretch of 7\% is needed. Thus, for each wavelength setting, we made a linear two-component adjustment of the wavelength scale (by applying wavelength shift and stretching) in order to match the reference spectrum. The adjustment process was done by computing and minimizing $\chi^2$. This gave an offset of $\sim$~0.01--0.02~$\mu$m (depending on the AMBER wavelength setting) with respect to the original wavelength table and a 0.01--0.02\% wavelength stretch. The latter is very small and can be neglected, as was done in many other studies of high-resolution AMBER observations \citep[e.g.][]{2009A&A...507..301D,2009A&A...503..183O}. This procedure also allows to set the velocity scale at 0 km/s\footnote{The uncertainty on our wavelength calibration is defined as the maximum wavelength shift which can be applied to the wavelength-corrected AMBER spectra in order to keep a good match with the combination of MARCS and telluric spectra. It amounts to about 7 km/s.}. This is a necessary step to perform the tomographic analysis described in Sect.~\ref{Sect:UD_fit} (since the tomographic masks are constructed from a static model atmosphere). This procedure was performed separately for the science and calibrator spectra. Then we applied the same calibration to the corresponding visibilities.  

Thereafter, the wavelength-corrected squared visibilities were calibrated with the instrumental transfer function derived from the calibrator. The errors on the calibrated squared visibilities were derived as follows. First, there is a systematic error associated with the percentage of the best frames being averaged. This error is derived as the difference between the average squared visibilities when 80\% and 20\% of the available frames are merged. It will be used as a proxy for the error on the transfer function, which could not otherwise be derived since only one calibrator star was observed. Other error terms on the calibrated squared visibilities arise from the  uncertainty on the measured angular diameter of the calibrator and from the statistical error inherent to the  averaging of single frames \citep[this error is provided by \textit{amdlib}, see Eq.~27 in][]{2007A&A...464...29T}. 

All visibility data obtained as described above are shown in Appendix~\ref{app:figures}, together with model predictions as outlined in Sect.~\ref{Sect:UD_fit}.

It is important to secure at close-enough epochs the different wavelength ranges necessary for the tomographic method to work best, in order to be able to combine them into a single snapshot, avoiding the disturbances caused by temporal variations.
Here, the retained data cover the narrow phase range 0.62 - 0.67. From their study of the Mira variable R~Peg, \citet{2018A&A...613L...7W} showed that the angular diameter of this Mira changed by $0.2\pm0.1$~mas only over this phase range. This variation is much smaller than the angular separation probed by two contiguous masks (as it will be shown in Table~\ref{tab:Uddiameters}). All the data in the final set listed in Table~\ref{tab:data_overview} may thus be used in the tomographic method and considered as representing a single epoch. Note that future data using GRAVITY will not face this difficulty any more, because this instrument covers the full $K$~band at once.

\section{Tomographic method}
\label{Sect:tomography}

The tomographic technique allows us to probe different optical depths in the stellar atmosphere and to recover the corresponding disk-averaged velocity field \citep[][and Paper~\citetalias{2018A&A...610A..29K}]{2001A&A...379..288A}. The implementation of the tomographic method described in Paper~\citetalias{2018A&A...610A..29K} was successfully applied to derive the line-of-sight velocity distribution at different optical depths in the atmosphere of the red supergiant star $\mu$~Cep (Paper~\citetalias{2019A&A...632A..28K}). The method is based on identifying the formation depth of different spectral lines, which is expressed in an optical depth scale computed at a reference wavelength of $\lambda = \rm 5000$~{\AA}. The formation depth is provided by the maximum of the line contribution function (CF, see Paper~\citetalias{2018A&A...610A..29K}). In doing so, we assign spectral lines to different masks according to the optical depths at which they form. 

   \begin{figure}
   \centering
    \includegraphics[width=8.9cm]{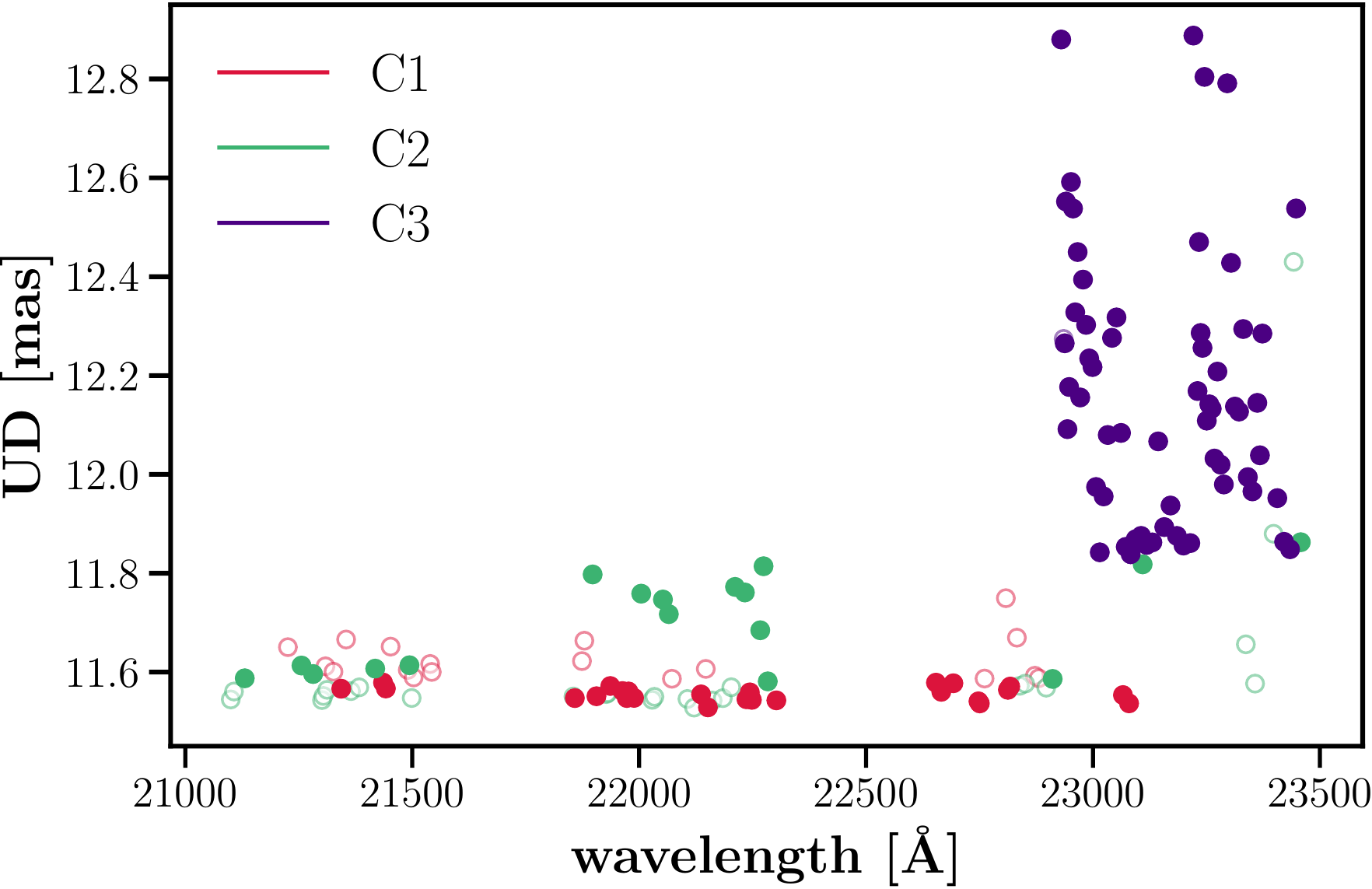}
      \caption{Angular diameters obtained from the fit of MARCS visibilities with a UD model, separately for each line of a mask. Lines contributing to the final set of masks constructed in Sect.~\ref{Sect:tomography} (see also Fig.~\ref{Fig:depth_function} and Table~\ref{tab:AMBERmasks}) are shown as filled circles. Open circles correspond to the lines removed from the masks due to mixing of their UD diameters.  
      }
         \label{Fig:MARCS_line_selecion}
   \end{figure}

   \begin{figure}
   \centering
    \includegraphics[width=9cm]{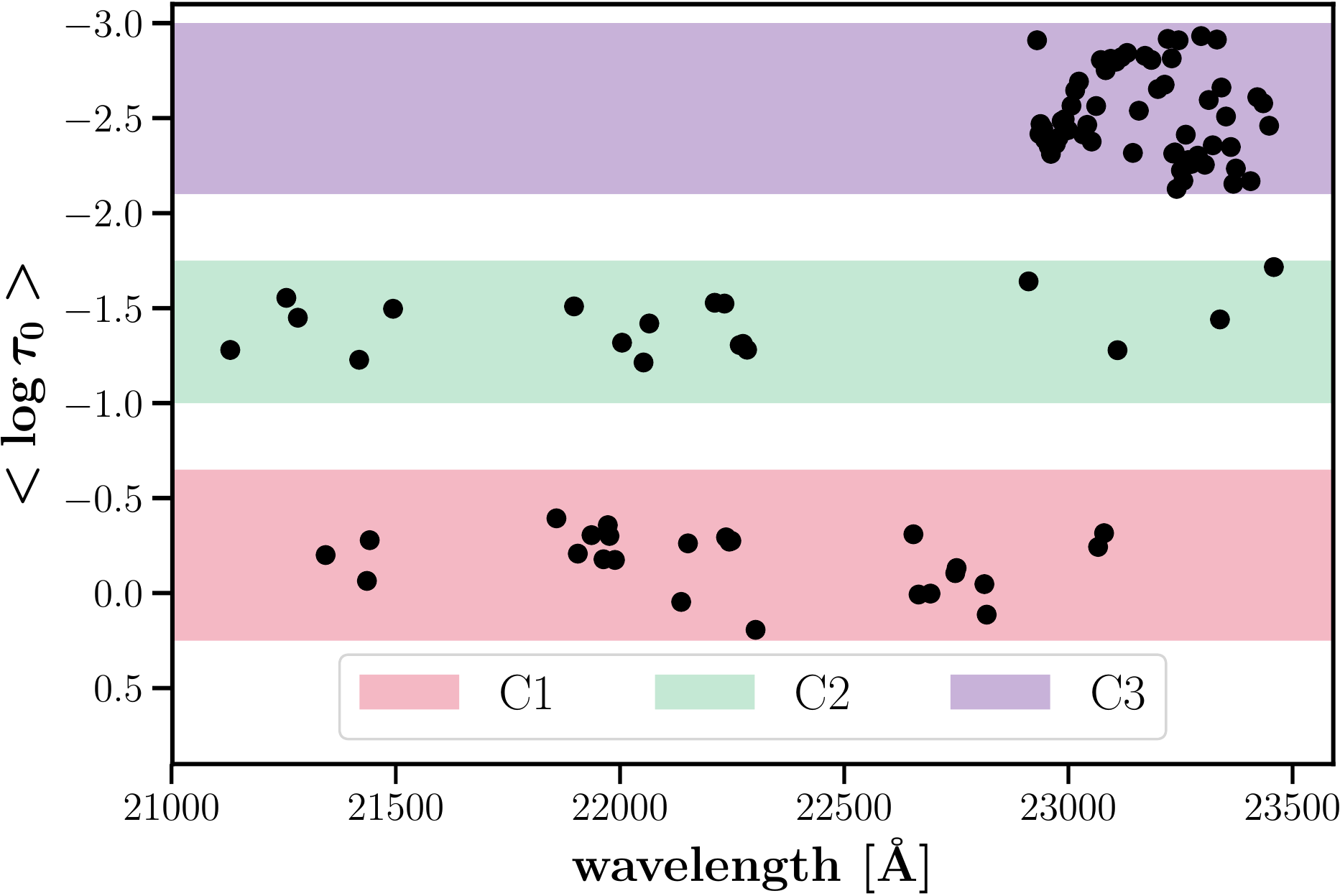}
      \caption{The average optical depth of lines contributing to the tomographic masks C1, C2 and C3 constructed in Sect.~\ref{Sect:tomography}. The horizontal bands mark the optical depth limits of the masks (see Table~\ref{tab:AMBERmasks}). 
      }
         \label{Fig:depth_function}
   \end{figure}

Following Papers~\citetalias{2018A&A...610A..29K} and \citetalias{2019A&A...632A..28K}, we designed masks from a 1D static model atmosphere. Paper~\citetalias{2018A&A...610A..29K} showed that atomic spectral lines defining a given mask from a static atmosphere form at similar optical depths in a dynamical atmosphere. This is true even when multiple layers contribute to a given line due to non-homogeneous temperature and density runs in a dynamical atmosphere. Thus, the tomographic method does not lose its discriminating power when applied to dynamical atmospheres (at least while considering only atomic lines and spectral resolution of $\sim$~100~000~--~200~000). For the present work, a set of three tomographic masks was constructed. The lines belong to spectral windows between 2.10~--~2.15 $\mu \rm{m}$, 2.18~--~2.24 $\mu \rm{m}$ and 2.26~--~2.35 $\mu \rm{m}$, i.e. those covered by our AMBER observations. For the mask construction, we used a 1D MARCS model atmosphere with stellar parameters ($T_{\rm eff} = 3000$~K, $\log g = 0.0$, and 1~$\rm M_{\odot}$) close to those of S Ori.

The mask construction process used  in the present study differs from that described in Papers~\citetalias{2018A&A...610A..29K} and \citetalias{2019A&A...632A..28K}. The spectral resolution of the AMBER instrument is about $12\,000$, and the computation of the CF from synthetic spectra at such a low resolution 
%(compared to high-resolution spectroscopy) 
would not lead to a reliable opacity evaluation when computing radiative transfer (since not all the opacities contributing to the lines inside each AMBER bin would be taken into account by the opacity sampling procedure used by the MARCS tools). Instead, our new approach is based on the derivation of the average depth of formation of the line depression \citep[Eq.~22 of][]{1986A&A...163..135M} from the high-resolution ($R \sim 200\,000$) CF and its subsequent averaging over low-resolution ($R \sim 12\,000$) wavelength bins:

\begin{equation}
    <\log \tau_0> = \frac{\int_{\lambda_i}^{\lambda_f} \rm{d}\lambda \int_{-\infty}^{+\infty} \rm{d} \log \tau_0 \log \tau_0 \; CF(\log \tau_0,\lambda) }{\int_{\lambda_i}^{\lambda_f} \rm{d}\lambda \int_{-\infty}^{+\infty} \rm{d} \log \tau_0 \; CF(\log \tau_0,\lambda) },
\label{Eq.1}    
\end{equation}
\
where $\lambda_i$ and $\lambda_f$ are the wavelength limits of low-resolution AMBER bins. This ensures that the contributions of the various spectral lines, which are blended at the spectral resolution of AMBER, are correctly taken into account. This, in turn, results in the inclusion of molecular lines in our masks.

 %--------------------------------------------------- One column table
\begin{table}
\begin{center}
\begin{threeparttable}
\caption[]{Properties of the tomographic masks constructed in Sect.~\ref{Sect:tomography}.}
\label{tab:AMBERmasks} 
%$$ 
\begin{tabular}{c c c }
\hline \hline
\noalign{\smallskip}
Mask  & $\rm \log \tau_0$ limits\tnote{*} & Number of lines   \\
\noalign{\smallskip}
\hline
\noalign{\smallskip}
C1 & $ -0.65 < \log \tau_0 \leq 0.25 $ & 25 \\
C2 & $ -1.75 < \log \tau_0 \leq -1.00 $ & 18 \\
C3 & $ -3.00 < \log \tau_0 \leq -2.10 $ & 61 \\
\noalign{\smallskip}
\hline
\end{tabular}
\begin{tablenotes}
\item [*] $\tau_0$ is the reference optical depth computed at $\lambda~=~5000$~\AA. 
\end{tablenotes}
\end{threeparttable}
\end{center}
\end{table}
%---------------------------------------------------

In our previous studies described in Papers~\citetalias{2018A&A...610A..29K} and \citetalias{2019A&A...632A..28K}, the tomographic masks have been constructed using high-resolution synthetic spectra from 1D static model atmospheres, keeping only atomic lines in masks. The high spectral resolution allowed us to discard blended spectral lines, thus avoiding to mix lines with different formation depths. The inclusion of atomic lines only, characterized by narrow CF profiles, allowed us to build masks probing non-overlapping optical-depth zones. However, in the present study, even the small bin size of AMBER is too wide to separate atomic and molecular lines (except within the CO band). Molecular lines are characterized by wide CF profiles and thus extend over a larger range of optical depths than atomic lines. In addition, while it was shown in Paper~\citetalias{2018A&A...610A..29K} that the masks containing only atomic lines do not lose their discriminating power when applied to dynamical atmospheres, it is not clear whether this remains the case for molecular lines. To summarize, by averaging the information pertaining to both atomic and molecular lines that probe different optical-depth ranges in the atmosphere, the resulting low-resolution masks lose their accuracy at probing the expected optical-depth range.

%--------------------------------------------------- One column table
\begin{table*}[h]
\begin{center}
\begin{threeparttable}
\caption[]{Parameters (variability phase $\Phi$, effective temperature $T_{\rm eff}$, surface gravity $\log g$,  luminosity $L$) and the photospheric angular diameter $\Theta_{\rm Phot}$ (derived from a fit of AMBER data with an atmosphere model) of 1D MARCS model (used for the masks construction in Sect.~\ref{Sect:tomography}), the best-fit 1D CODEX model and the best-fit snapshot from the 3D RHD CO5BOLD simulation used in the present work.}
\label{tab:stellar_parameters} 
%$$ 
\begin{tabular}{c c c c c c}
\hline \hline
\noalign{\smallskip}
 Model & $\Phi$ & $T_{\rm eff}$ & $\log g$   &  $L$ & $\Theta_{\rm Phot}$  \\ %& Ref. \\
   &  & [K]   &  [c.g.s.]    &    [$\rm L_{\odot}$] & [mas]   \\
\noalign{\smallskip}
\hline
\noalign{\smallskip}
1D MARCS &  & 3000 & 0.00 &  2026 & 11.79 $\pm$ 0.02  \\
 & &  & & &  \\
1D CODEX $o$54/289240 & 0.3 & 2948 & -0.48\tnote{a} & 6174 & 10.50 $\pm$ 0.05  \\
 & & &  & &  \\
3D CO5BOLD st29gm06n001/parambf002 &  & 2885\tnote{b} & -0.66\tnote{b} &  7790 & 11.45 $\pm$ 0.03  \\  
\noalign{\smallskip}
\hline
\end{tabular}
\begin{tablenotes}
\item [a] Derived from the mass and the radius of the model.
\item [b] Averages over spherical shells \citep{2009A&A...506.1351C}.
\end{tablenotes}
\end{threeparttable}
\end{center}
\end{table*}
%---------------------------------------------------

Thus, we checked the ability of each mask line to probe the expected optical-depth range by deriving the corresponding UD diameter from the MARCS model. As a first step, we computed synthetic visibilities at the spectral resolution of AMBER from the tabulated intensity profiles for the MARCS model using the Hankel transform. Then, we extracted squared visibility amplitudes at wavelengths contributing to the tomographic masks and fitted them with a UD model (about the choice of a UD model, see our discussion in Sect.~\ref{Sect:UD_fit}). The resulting UD angular diameters are shown in Fig.~\ref{Fig:MARCS_line_selecion} as open and filled circles. They overlap between masks C1 and C2 as consequence of inclusion of molecular lines in the masks. Thus, the final set of masks was constructed by keeping only lines whose UD diameters are well separated between the masks C1 and C2 (see filled circles in Fig.~\ref{Fig:MARCS_line_selecion}). In so doing, our masks are separated both in terms of optical and geometrical depths. Figure~\ref{Fig:depth_function} shows the average optical depths derived with Eq.~\ref{Eq.1} for the adopted lines contributing to the optimized tomographic masks. The optical-depth ranges and the number of lines in the optimized masks are summarized in Table~\ref{tab:AMBERmasks}.

Our choice to build only three masks is partially driven by the need to include a significant number of lines per mask. Moreover, interferometric studies of evolved-star atmospheres generally reveal the presence of at least two atmospheric layers, one corresponding to the near-continuum \citep[used to measure the photospheric angular diameters of stars; e.g.,][]{2009A&A...503..183O,2017A&A...597A...9W} and the other to the CO-band formation region \citep[used to measure the star's  atmospheric extension; e.g.][]{2015A&A...575A..50A}. However, recent studies of the atmosphere of AGB stars showed the presence of an intermediate layer containing weak molecular and atomic lines \citep[e.g.,][]{2016A&A...589A..91O,2019A&A...621A...6O}. Indeed, in Fig.~\ref{Fig:depth_function}, the near-continuum (characterized by the largest optical depths) and CO-band (characterized by the lowest optical depths and located at wavelengths larger than 22800 \AA) regions are clearly distinguishable. These regions are probed by masks C1 and C3, respectively. Mask C2, in turn, collects lines falling in the intermediate region between masks C1 and C3.

In order to investigate the properties of the tomographic masks, we identified the dominant atomic and molecular features contributing to the masks. Tables~\ref{tab:C1}, \ref{tab:C2}, and \ref{tab:C3} summarize the properties (element, central wavelength, excitation potential, $\rm \log gf$ and equivalent width) of spectral lines probed by masks C1, C2, and C3, respectively. We found that mask C3 contains only CO lines. Mask C2 is characterized by mostly atomic (\ion{Ti}{I} and \ion{Sc}{I}) and $\rm H_2O$ lines, as well as two CO lines. Mask C1 contains mostly CN lines, a few atomic lines and one OH line. The difference in chemical properties of the lines contributing to the tomographic masks further supports our choice to build only three masks.

\section{Calibration of the tomographic masks in terms of geometrical depth}
\label{Sect:UD_fit}

In this Section, we aim at calibrating the tomographic masks probing different optical depths in terms of geometrical depths, taking S~Ori as benchmark. This can be done using interferometry, by measuring the stellar diameter at the wavelengths probed by each given mask. For this purpose,  visibilities  must be estimated at wavelengths contributing to the tomographic masks, and fitted to a geometrical model to convert them into angular diameters. Combined with the stellar parallax, these angular diameters yield in turn  the geometrical extension of the atmospheric layer probed by each mask.

In the fitting process, we considered only AMBER visibilities obtained with the small (G0-E0) and medium (H0-G0) baselines. Visibilities from the large baseline (E0-H0, $V^2$~<~0.1) are not used since they correspond to spatial frequencies close to the first zero of the visibility function, where higher-order effects linked to the surface inhomogeneities strongly contribute \citep{2011A&A...532L...7W}. 

\subsection{S Ori: UD-model fit}
\label{UD_fit}

We applied the above procedure to the Mira star S~Ori (Sect.~\ref{Sect:observations}), and a UD model has been used. This choice is imposed by the small number of available visibility points in the Fourier $uv$ plane (see Fig.~\ref{Fig:uv_plot}), which makes it impossible to reconstruct the complete (limb-darkened) intensity profile of S~Ori and, thus, to constrain more complicated intensity distribution models. Since the surface-brightness distribution of a Mira star differs from a UD \citep{2008A&A...479L..21W}, a UD fit will provide an estimate of the {\it relative} spatial extensions of the different line-forming regions. In any case, this procedure is adequate to match our primary goal which is to check whether the tomographic masks probe distinct atmospheric layers. It allows us to relate optical- and geometrical-depth scales. The accurate measurement of the atmospheric extension is deferred to a forthcoming paper using VLTI/GRAVITY \citep{2017A&A...602A..94G} data.

\begin{figure}[h!]
   \centering
   \includegraphics[width=9cm]{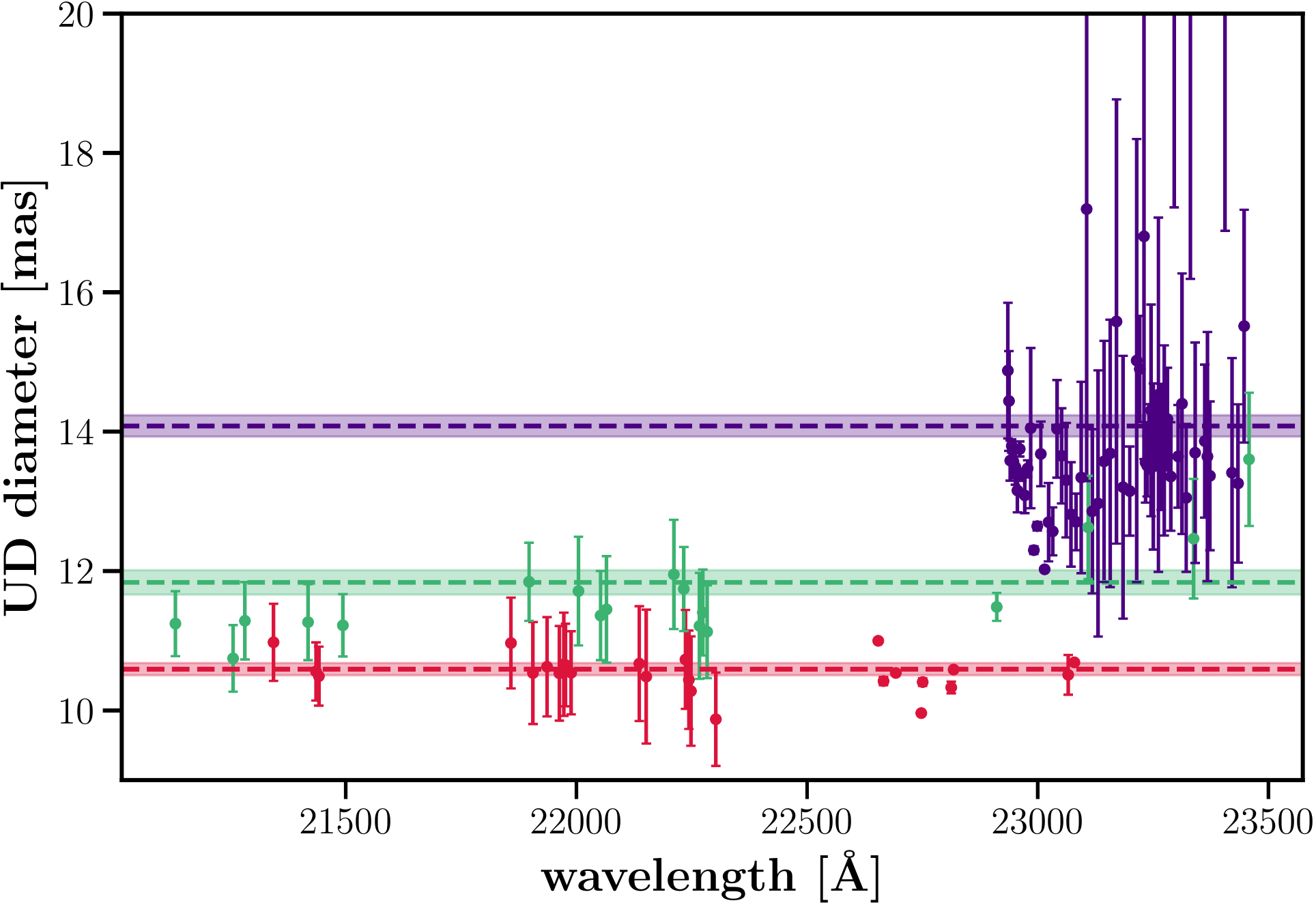}\\
   \includegraphics[width=9cm]{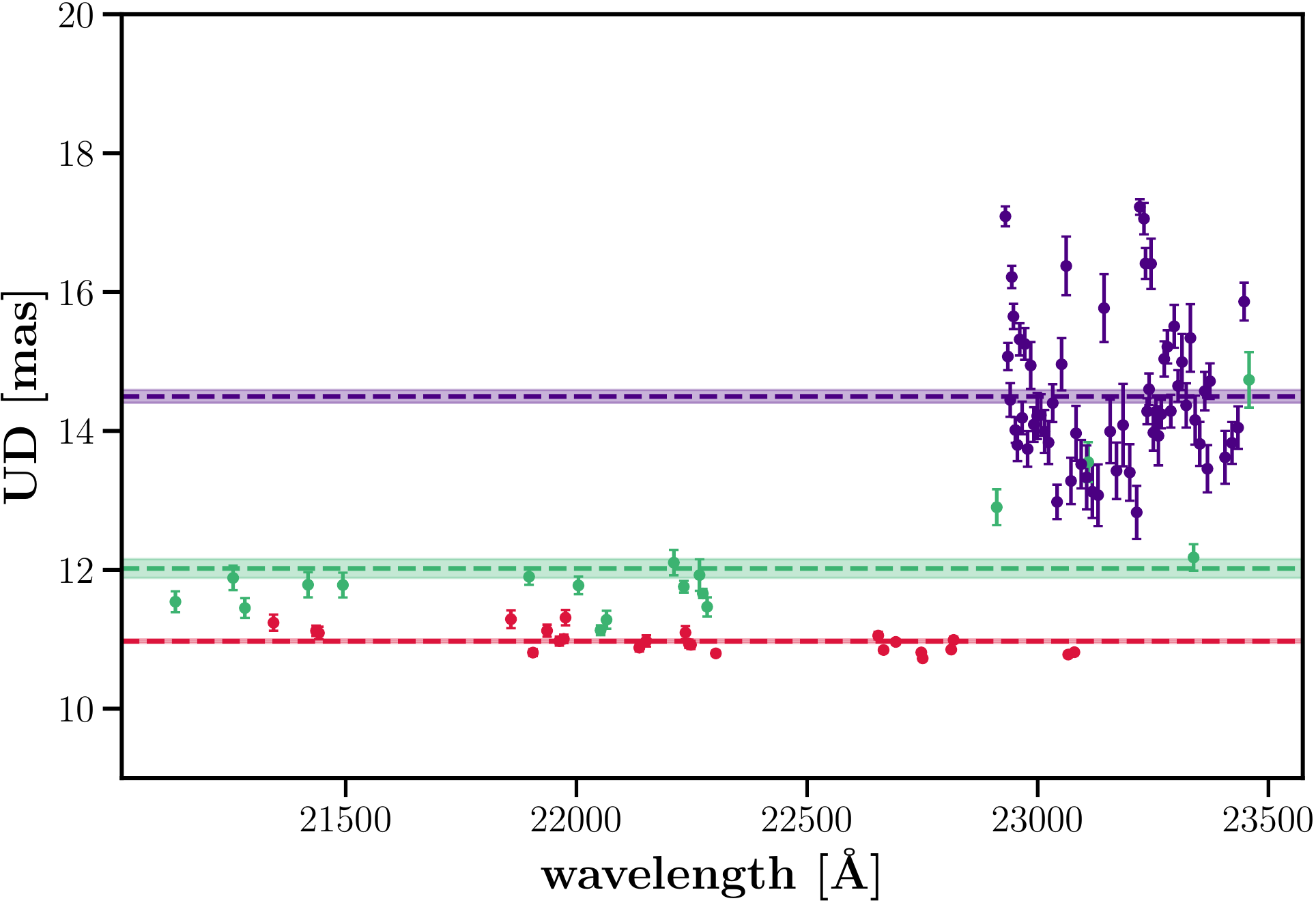}\\
   \includegraphics[width=9cm]{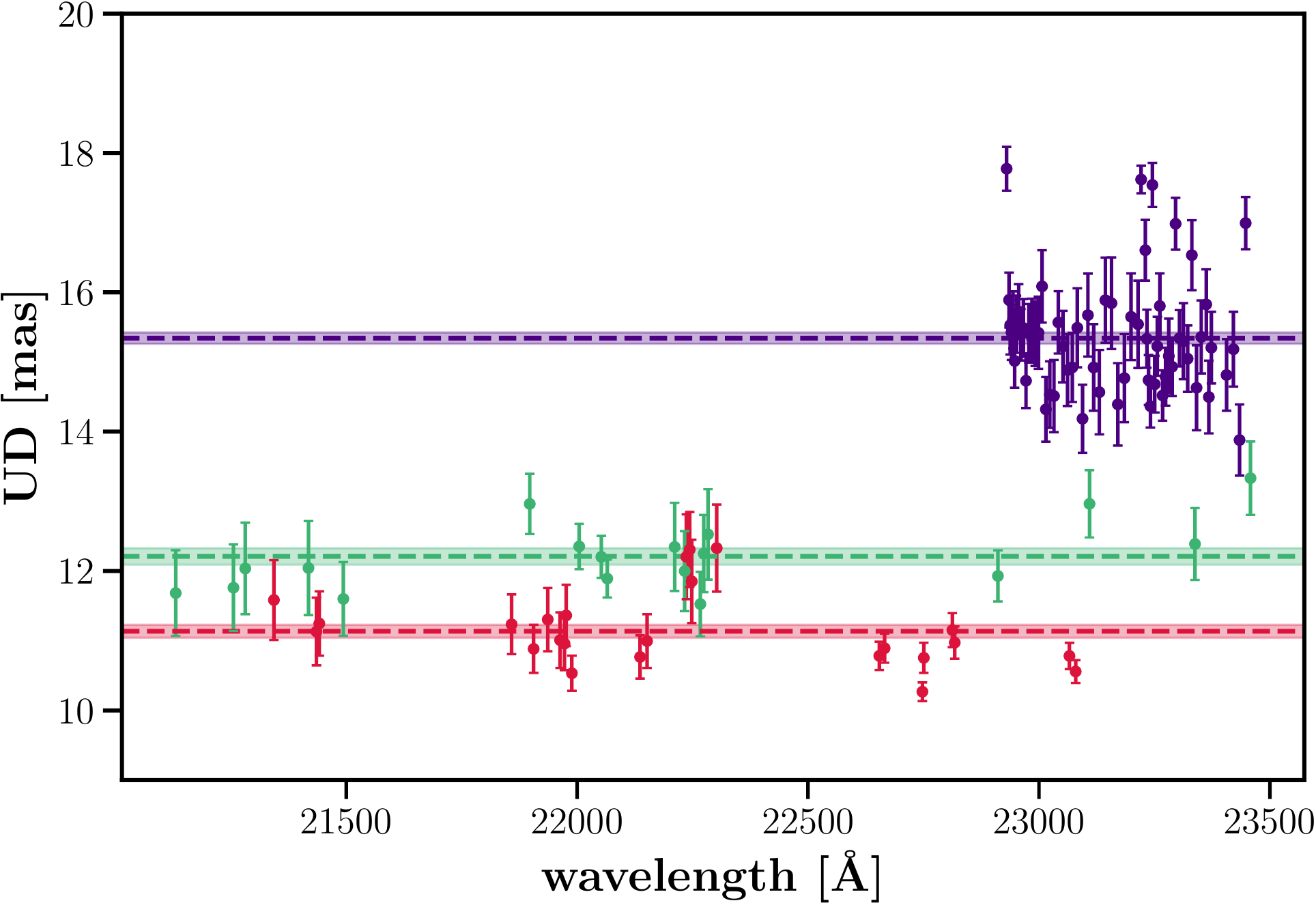}   
      \caption{\textit{Top panel:} Angular diameters obtained from the fit of AMBER visibilities with a UD model, separately for each line of a mask. Lines belonging to a given mask are coded in the corresponding color (see Fig.~\ref{Fig:depth_function}).  Dashed horizontal lines correspond to the UD diameters obtained from a UD-model fit of all AMBER visibilities contributing to a given mask at once. Shaded areas correspond to errorbars. \textit{Middle panel:} Same as top panel for the best-matching 1D CODEX model (see Sect.~\ref{CODEX_UD}). \textit{Bottom panel:} Same as top panel for the best-matching snapshot from the 3D RHD simulation (see Sect.~\ref{3D_UD}).
      }
         \label{Fig:UDfit}
   \end{figure}

The UD fit was performed in two ways. First, we fitted the squared visibility amplitudes separately for each line in a given mask. The resulting UD angular diameters are shown as dots with error bars in the top panel of Fig.~\ref{Fig:UDfit}. Second, we performed a UD-model fit of visibilities for \textit{all} lines contributing to a given mask at once (see Fig.~\ref{Fig:AMBER_spatfreq_plot} illustrating the fit). The resulting angular diameters for this global approach show a steady increase from mask C1 to C3 with values of 10.59 $\pm$ 0.09, 11.84 $\pm$ 0.17, and 14.08 $\pm$ 0.15 mas (see Table~\ref{tab:Uddiameters}) and are shown in the top panel of Fig.~\ref{Fig:UDfit} as horizontal dashed lines. Results from the two methods agree fairly well and reveal a clear increase of the angular diameter with increasing mask number or decreasing reference optical depth. Thus, Fig.~\ref{Fig:UDfit} confirms the capability of the tomographic method to probe distinct geometrical depths in the stellar atmosphere. Finally, Fig.~\ref{Fig:UD_vs_logtauref} displays our measured UD diameters as a function of the reference optical depth for each line of a tomographic masks (top panel) and averaged over a given mask (bottom panel). Thus, Fig.~\ref{Fig:UD_vs_logtauref} achieves our goal of linking geometrical- and optical-depth scales. 

The UD diameters of S Ori derived in the top panel of Fig.~\ref{Fig:UDfit} differ from those derived in previous studies. For example, compared to the UD diameters derived from low-resolution AMBER observations of S Ori by \citet{2008A&A...479L..21W} at phase 0.78 (9.5~mas in the $K$-band near-continuum, and 10~mas in the CO band), our UD diameters in the near-continuum (mask C1) and CO band (mask C3) are larger. This could result from the fact that, with the tomographic method, angular diameters are measured in specific spectral lines rather than averaging visibilities over spectral regions containing both continuum and lines. In addition, the systematic uncertainties in the absolute visibility calibration of S~Ori (Sect.~\ref{Sect:observations}) may partially be responsible for the discrepancy.

We converted the UD angular diameters of S~Ori into linear radii using a distance based on the period-luminosity (P-L) relation. \citet{2008MNRAS.386..313W} derived the $K$-band P-L relation for O-rich Galactic Miras, which resulted in a distance of 390~pc to S~Ori. Using this value for the distance, we computed the linear radii of S~Ori in the different masks. We found that the radius of S~Ori is $2.07 \pm 0.02$~au in mask C1, $2.31 \pm 0.03$~au in mask C2, and $2.75 \pm 0.03$~au in mask C3. These values are consistent with previous observations of AGB stars. In particular, the radius of S~Ori in mask C1 is comparable to the photospheric radii of Mira variables from \citet{2016A&A...587A..12W}. In addition, the atmospheric extension of S~Ori in mask C3 is consistent with the previous observations of other AGB stars by \citet{2016A&A...589A..91O,2019A&A...621A...6O}, who showed that their molecular layers are located at $\sim$1.5~$R_{*}$ and further away. A similar agreement is present in the results from the dynamical model atmospheres reported in Sects.~\ref{CODEX_UD} and \ref{3D_UD} (see also Table~\ref{tab:Uddiameters}).

Another way to convert the UD angular diameters into linear radii is by using the Gaia DR2 parallax. Unfortunately, nowadays astrometric observations of Mira stars experience problems which strongly affect the accuracy of stellar parallaxes. This is particularly the case for Miras with diameters larger than their parallaxes. First, Mira stars are characterized by very red colors and, thus, produce chromatic effects caused by the wavelength-dependent diffraction. Second, the variability of Mira stars leads to colour variations over the light cycle. Third, the convection-related variability in these stars produces a displacement of the photocentre with respect to the barycentre. In particular, \citet{2018A&A...617L...1C} analyzed 3D RHD simulations of AGB stars and showed that the position of the photocentre moves by $\sim$5--11\% of the stellar radius and accounts for a significant fraction of the Gaia DR2 parallax error. Finally, the photocenter position and the surface pattern of Mira stars change with time scales similar to the parallactic motion. Improvements should come with the future Gaia data releases. Since it is not possible to use the Gaia DR2 parallax to compare with the results from the P-L relation and since the comparison of linear sizes between S~Ori and dynamical models is meaningless for the same reasons, the following sections only describe the comparison of angular diameters.

\subsection{Comparison with 1D pulsation CODEX models}
\label{CODEX_UD}

In this Section, we compare AMBER visibilities of S Ori to those predicted by 1D self-excited pulsation CODEX model atmospheres of Mira variables \citep{2008MNRAS.391.1994I,2011MNRAS.418..114I}. We selected the $o54$ model series with $T_{\rm eff} = 3370$~K (the closest to S~Ori as compared to other model series\footnote{There are only four CODEX model series available, and none of them matches well the parameters of S Ori.}), mass of 1.1~M$_{\odot}$ and pulsation period of 330~d. The model series consists of 66 snapshots, each of them corresponding to a different variability phase. As a first step, we computed synthetic visibilities from the tabulated intensity profiles \citep[as in][]{2016A&A...587A..12W} for all the snapshots from the model series using the Hankel transform. To account for the spectral resolution of AMBER, we averaged the monochromatic squared visibility amplitudes over the AMBER wavelength bins. Then, we fitted the S~Ori visibilities to every snapshot from the model series by keeping the photospheric angular diameter $\Theta_{\rm Phot}$\footnote{In CODEX models, the photospheric radius $R_{\rm Phot}$ and photospheric angular diameter $\Theta_{\rm Phot}$ are defined as the Rosseland linear radius ($R_{\rm Ross}$) and the Rosseland angular diameter ($\Theta_{\rm Ross}$), respectively. $R_{\rm Ross}$ and $\Theta_{\rm Ross}$ correspond to the model layer where the Rosseland optical depth equals unity \citep{2008MNRAS.391.1994I,2011MNRAS.418..114I}.} as a free parameter. The fitting process and the selection of the best-fit snapshot was done by computing and minimizing $\chi^2$, where the error term corresponds to the statistical errors on the squared visibility amplitudes of AMBER observations (see Sect.~\ref{Sect:observations}). Table~\ref{tab:stellar_parameters} summarizes the parameters of the best-matching CODEX snapshot $o$54/289240, which, in addition, has parameters close to those of S~Ori. The reduced $\chi^2$ value equals 84. The next best $\chi^2$ values are above 90. The large $\chi^2$ may arise from uncertainties in the absolute visibility calibration of S~Ori or/and mismatch in the model stratification with respect to S~Ori. This has little impact on our analysis since we aim at comparing the relative spatial extensions in different tomographic masks, for S~Ori and for dynamical models, rather than measuring accurate stellar diameters. The synthetic visibilities of the best-fit CODEX snapshot are compared to those of S~Ori in Appendix~\ref{app:figures} for all AMBER wavelength settings and reveal a good agreement with the observations. In particular, the snapshot reproduces the observed decrease in the visibility function at the locations of the CO bands. This further confirms the capability of 1D CODEX models to predict extended molecular atmospheres of Mira stars \citep[as already shown by][for other Mira variables]{2016A&A...587A..12W}.

In order to check whether the CODEX model atmospheres predict larger UD diameters with increasing mask number as well, we extracted squared visibility amplitudes computed from our selected CODEX snapshot at wavelengths contributing to the tomographic masks and fitted them with a UD model in the same way as was done in Sect.~\ref{UD_fit} for S~Ori.

Results are shown in the middle panel of Fig.~\ref{Fig:UDfit}, in Table~\ref{tab:Uddiameters}, and in Fig.~\ref{Fig:CODEX_spatfreq_plot}. As for S~Ori, the UD diameters of the CODEX snapshot increase along the mask sequence  C1, C2 to C3 with values 10.97 $\pm$ 0.09, 12.02 $\pm$ 0.14, and 14.50 $\pm$ 0.09~mas, respectively. Thus, the relation between geometrical and optical depth scales illustrated in Fig.~\ref{Fig:UD_vs_logtauref} is confirmed by predictions from 1D pulsation CODEX model atmospheres.

\subsection{Comparison with 3D radiative-hydrodynamics CO5BOLD simulations}
\label{3D_UD}

In this Section, we compare AMBER visibilities of S~Ori to those predicted by time-dependent 3D radiative-hydrodynamics (RHD) simulations of convection in the outer envelope and atmosphere of AGB stars, as computed by \citet{2017A&A...600A.137F} using the CO5BOLD code \citep{2012JCoPh.231..919F}. The model geometry is of the kind "star-in-a-box". It is characterized by an equidistant cartesian grid with the same open boundary conditions for all sides of the simulation box. 

From the sample of \citet{2017A&A...600A.137F}, we selected the st29gm06n001 simulation  having stellar parameters closely matching those of S~Ori: $T_{\rm eff} = 2822$~K, $\log g = -0.65$, mass 1~M$_{\odot}$ and pulsation period of 1.15~yr (i.e. about 420~d). The simulation is characterized by the "grey" frequency dependence of the radiation field. The pulsations are self-excited, as in the 1D~CODEX models, and the pulsation period and amplitude are outputs of the 3D simulation.

We calculated intensity maps for 13 snapshots of the simulation using the pure-LTE radiative-transfer code Optim3D \citep{2009A&A...506.1351C}. The code performs the detailed radiative transfer using opacity tables constructed as a function of temperature and density, and using solar elemental abundances from \citet{2009ARA&A..47..481A}. Instead of micro-turbulence broadening, Optim3D takes into account the Doppler shifts caused by the convective motions in the stellar atmosphere. The intensity maps were computed for the spectral windows 2.10~--~2.15 $\mu \rm{m}$, 2.18~--~2.24 $\mu \rm{m}$ and 2.26~--~2.35 $\mu \rm{m}$ to cover the wavelength range of our AMBER observations. We used a constant spectral resolution of 60~000 in order to average in a subsequent step the monochromatic squared visibilities over each AMBER spectral channel. Next, we computed the intensity profiles by azimuthally averaging the intensity maps \citep[as explained in][]{2009A&A...506.1351C}. Then, the synthetic visibilities were computed from the intensity profiles using the Hankel transform and averaging over each AMBER spectral bin. 

The following steps were performed in the same way as for the CODEX model series described in Sect.~\ref{CODEX_UD}. First, we fitted visibilities of S Ori to every 3D snapshot until finding the best-matching one. The photospheric angular diameter $\Theta_{\rm Phot}$\footnote{In CO5BOLD simulations, the photospheric radius $R_{\rm Phot}$ of a snapshot corresponds to the snapshot's radius, which is calculated from the luminosity and effective temperature averaged over spherical shells \citep[see][for more details]{2009A&A...506.1351C}.} has been kept as a free parameter during the fit (see Sect.~\ref{CODEX_UD} for the fitting procedure). The parameters of the best-fit snapshot are shown in Table~\ref{tab:stellar_parameters}, and the reduced $\chi^2$ equals 51, and the next best $\chi^2$ are above 60 (about the large $\chi^2$ value even for the best-fit snapshot, see the discussion in Sect.~\ref{CODEX_UD}). Its synthetic visibilities are compared to those of S~Ori and to the best-fit 1D CODEX snapshot in Appendix~\ref{app:figures}, and are consistent with both. The 3D snapshot visibilities nicely reproduce the decrease of the S~Ori visibilities in the CO band. This means that, similar to CODEX models, 3D RHD simulations are also able to predict the extended molecular layers in Mira-star atmospheres \citep{2016A&A...587A..12W}. The photospheric diameters $\Theta_{\rm Phot}$ of the best-fit CODEX snapshot and of the 3D snapshot in Table~\ref{tab:stellar_parameters} are comparable, considering the differences in stellar parameters and in the definitions of $R_{\rm Phot}$ (and $\Theta_{\rm Phot}$)  between the models.

\begin{figure}
   \centering
   \includegraphics[width=8.5cm]{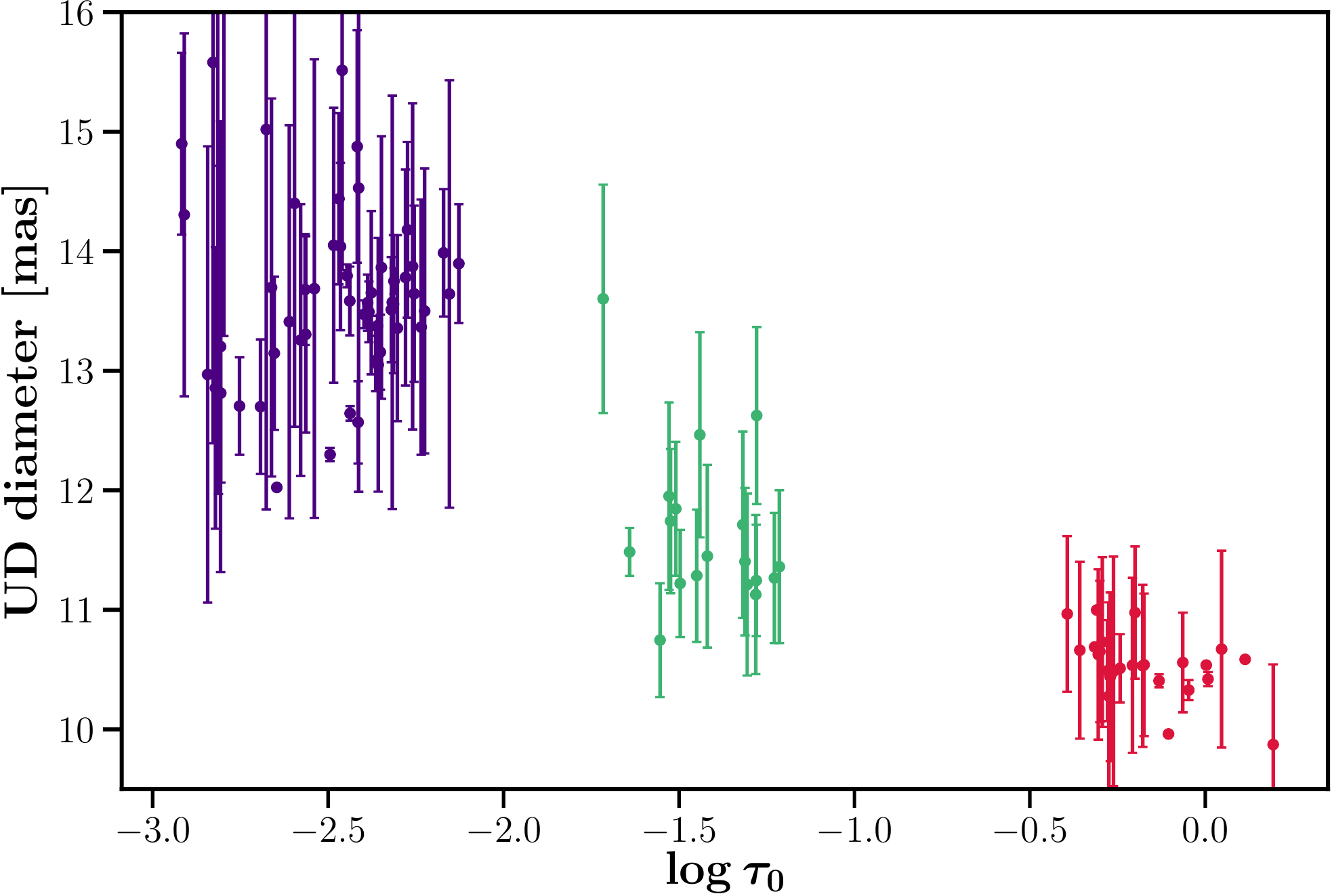}
   \includegraphics[width=8.6cm]{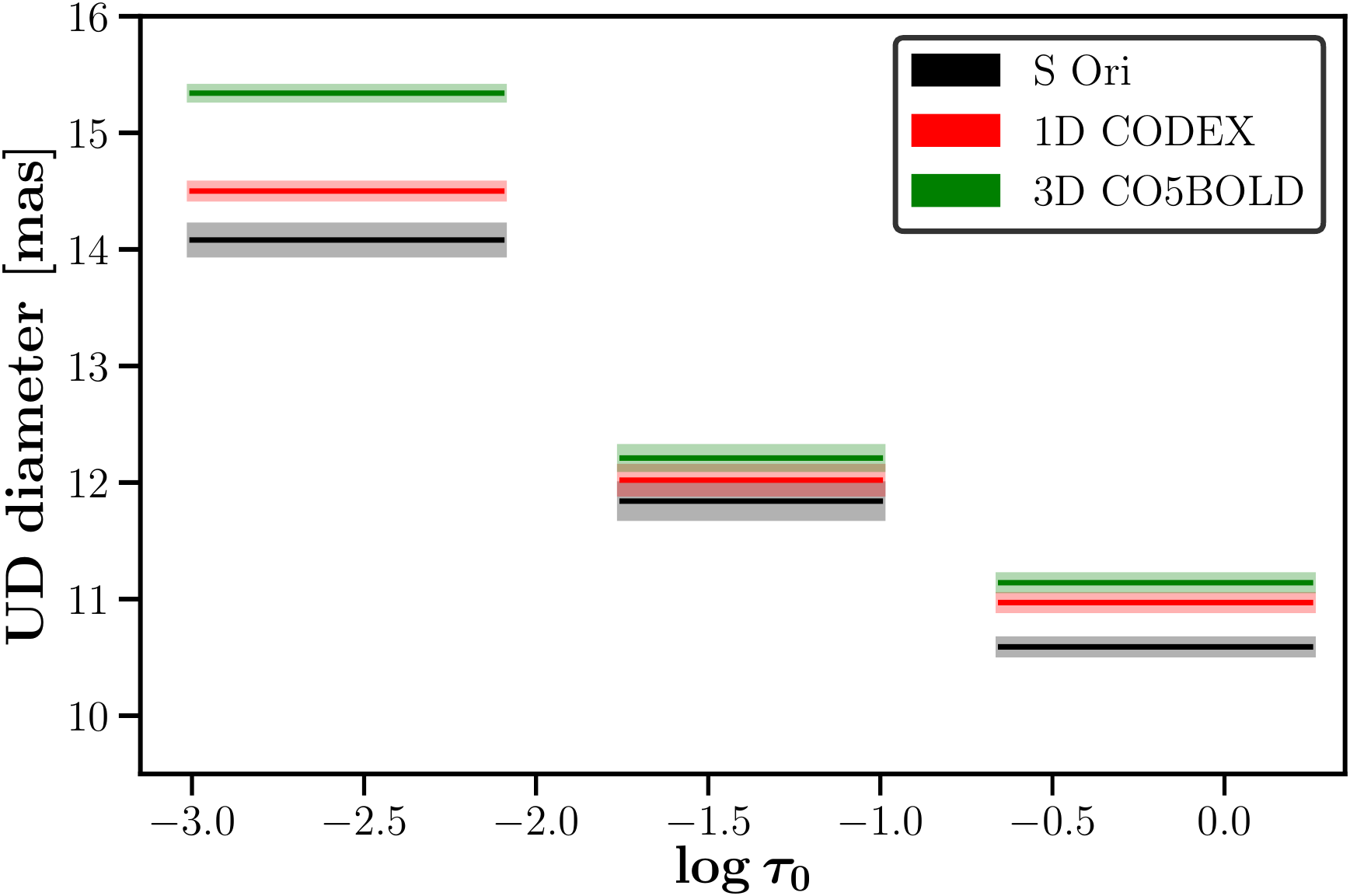}

      \caption{\textit{Top panel:} UD angular diameters of S~Ori for each line of a mask versus their corresponding reference optical depths (Fig.~\ref{Fig:depth_function}). The color coding is the same as in Figs.~\ref{Fig:depth_function} and \ref{Fig:UDfit}. \textit{Bottom panel:} UD angular diameters of S Ori (black), the best-fit 1D CODEX snapshot (red) and the best-fit snapshot from the 3D RHD simulation (green) from Table~\ref{tab:Uddiameters} for the different tomographic masks versus their corresponding reference optical depths (Table~\ref{tab:AMBERmasks}). The errorbars are shown as shaded areas. 
     }
         \label{Fig:UD_vs_logtauref}
          \end{figure}

As a final step, we extracted the squared visibility amplitudes of the best-matching 3D snapshot and performed a UD-model fit in two ways as was done in Sects.~\ref{UD_fit} and \ref{CODEX_UD}. The resulting UD diameters for different tomographic masks are shown in the bottom panel of Fig.~\ref{Fig:UDfit} and are compared to those of S Ori and to the best-fit 1D CODEX snapshot in Table~\ref{tab:Uddiameters} and Fig.~\ref{Fig:UD_vs_logtauref} (see Fig.~\ref{Fig:CO5BOLD_spatfreq_plot} illustrating the fit). The UD diameters of the 3D snapshot increase from mask C1 to mask C3 with values 11.13 $\pm$ 0.09, 12.21 $\pm$ 0.12, and $15.34\pm0.08$~mas. Thus, the link between optical and geometrical depth scales displayed in Fig.~\ref{Fig:UD_vs_logtauref} is further validated by 3D stellar convection simulations.

According to Table~\ref{tab:Uddiameters}, the UD diameters of the best 1D and 3D snapshots are systematically larger than those measured for S~Ori. As mentioned in the previous sections, there are systematic uncertainties in the absolute visibility calibration of S~Ori, which may contribute to the discrepancy. In addition, the stratification of the CODEX and CO5BOLD atmosphere models may not perfectly match S~Ori. Moreover, the stellar parameters of the models are not exactly the same as the observed ones. Nevertheless, the relative distances between the masks, and even absolute UD angular diameters, are very well reproduced by the models.

 %--------------------------------------------------- One column table
\begin{table}
\begin{center}
\caption[]{UD angular diameters of S Ori for the different tomographic masks constructed in Sect.~\ref{Sect:tomography} compared to those from the best-fit 1D pulsation CODEX model and the snapshot from the 3D RHD CO5BOLD simulation.} 
\label{tab:Uddiameters} 
%$$ 
\begin{tabular}{c c c c}
\hline \hline
\noalign{\smallskip}
Mask  & $\rm UD_{S \, Ori}$   & $\rm \Theta_{CODEX}$ & $\rm \Theta_{CO5BOLD}$  \\
  & [mas]   &  [mas] &   [mas]\\
\noalign{\smallskip}
\hline
\noalign{\smallskip}
C1 & 10.59 $\pm$ 0.09  & 10.97 $\pm$ 0.09 & 11.14 $\pm$ 0.09 \\
C2 & 11.84 $\pm$ 0.17  & 12.02 $\pm$ 0.14 & 12.21 $\pm$ 0.12  \\
C3 & 14.08 $\pm$ 0.15  & 14.50 $\pm$ 0.09 & 15.34 $\pm$ 0.08  \\
\noalign{\smallskip}
\hline
\end{tabular}
\end{center}
\end{table}

\section{Summary and conclusions}
\label{Sect:conclusions}

   \begin{figure}
   \centering
   \includegraphics[width=9.cm]{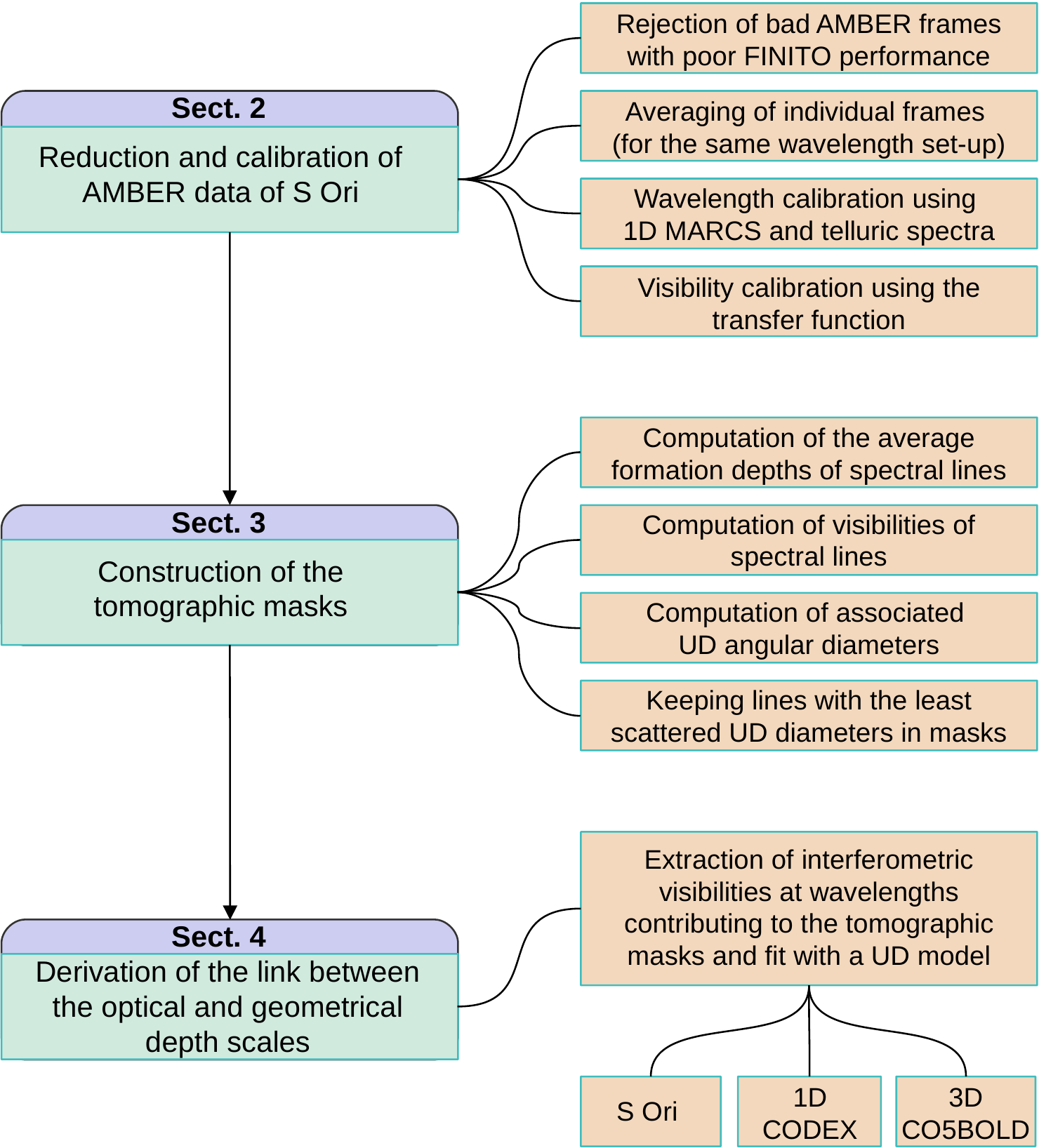}
      \caption{The flow chart of the methodology used in the paper.
      }
         \label{Fig:cookbook}
   \end{figure}

In the present paper, we report the results of the first application of the tomographic method to high-resolution near-IR spectro-interferometric VLTI/AMBER observations of the Mira star S~Ori. The flow chart of the methodology used in the paper is illustrated in Fig.~\ref{Fig:cookbook}. We extracted interferometric squared visibility amplitudes of S~Ori at wavelengths contributing to the tomographic masks and fitted them with a UD model in order to estimate the relative spatial extension of different line-forming regions. We observe an increase of the measured UD angular with decreasing optical depths probed by the tomographic masks. This reveals, for the first time, the relation between optical (provided by tomography) and geometrical (provided by interferometry) depth scales for a Mira-type star. Comparison to 1D pulsation CODEX model atmospheres and 3D RHD convection simulations of M-type AGB stars allowed us to recover similar relations, thus supporting the S~Ori results. The recovery of the link between optical and geometrical depth scales further validates the tomographic method and its capability to probe distinct atmospheric layers. 

In addition, we detected in S~Ori an intermediate atmospheric layer probed by the mask C2 (optical depth $-1.75<\log \tau_0 \leq -1.00$, UD diameter $11.84 \pm 0.17$ mas) located slightly above the near-continuum layer (mask C1, $-0.65<\log \tau_0 \leq 0.25$, $10.59 \pm 0.09$ mas) and much less extended than the CO layer (mask C3, $-3.00<\log \tau_0 \leq -2.10$, $14.08 \pm 0.15$~mas). The presence of such an intermediate layer in both AMBER observations and dynamical models as well as its chemical composition (mainly strong atomic lines and molecular $\rm H_2O$ lines) indicate that atmospheres of AGB stars have a very complex multi-layer structure unlike modeled by some earlier theories \citep[e.g. MOLsphere,][]{2000ApJ...540L..99T}.

In the future, we aim at expanding the present study to spectro-interferometric observations of Mira-type stars, and S~Ori in particular, with the VLTI/GRAVITY instrument. It is characterized by higher fringe-tracking precision and combines four telescopes, thus improving the $uv$ coverage. The latter is important in the context of recovering the true intensity profile of a star in order to constrain more complex geometrical models (rather than using UD) and to derive more accurate angular diameters in different tomographic masks. 

The knowledge of the link between optical and geometrical depth scales opens a way to derive the shock-wave propagation velocity in atmospheres of Mira-type stars, which has not been measured yet. This can be done by combining high-resolution spectro-interferometric and spectroscopic observations of Mira stars. This will be possible in the near future by combining VLTI/GRAVITY and VLT/CRIRES+ \citep{2014SPIE.9147E..19F} instruments or VLT/CRIRES+ and VLT/SPHERE \citep{2018SPIE10702E..36V} instruments in the near-IR domain. The knowledge of the shock wave velocity in real stars will allow us to perform a unique test of dynamical model atmospheres and to better understand the role of shocks in driving the mass loss. In addition, the derivation of the link between optical and geometrical depth scales for a sample of close benchmark stars will provide constraints to 1D and 3D dynamical model atmospheres, which will be later used to interpret more distant and only spectrally accessible stars. 

The tomographic method is a valuable tool, which has potential to simultaneously probe the structure and dynamics of stellar atmosphere. In future, the application of the tomographic method to different types of observations will allow us to obtain a three-dimensional view on stellar atmospheres and to better understand the processes and mechanisms driving the mass loss in evolved stars.

\begin{acknowledgements}
This research has made use of the \texttt{AMBER data reduction package}\footnote{\url{http://www.jmmc.fr/amberdrs}} and JSDC catalogue\footnote{\url{http://www.jmmc.fr/catalogue\_jsdc.htm}} of the Jean-Marie Mariotti Center (JMMC). This work has made use of data from the European Space Agency (ESA) mission {\it Gaia} (\url{https://www.cosmos.esa.int/gaia}), processed by the {\it Gaia} Data Processing and Analysis Consortium (DPAC, \url{https://www.cosmos.esa.int/web/gaia/dpac/consortium}). Funding for the DPAC has been provided by national institutions, in particular the institutions participating in the {\it Gaia} Multilateral Agreement. S.V.E. acknowledges support of Fondation ULB. B.F. acknowledges support from the Swedish Research Council (Vetenskapsr{\aa}det). The CO5BOLD model was computed on resources provided by SNIC through Uppsala Multidisciplinary Center for Advanced Computational Science (UPPMAX). We acknowledge with thanks the variable star observations from the AAVSO International Database contributed by observers worldwide and used in this research. This research has made use of the SIMBAD and AFOEV databases, operated at the CDS, France. This research made use of NASA’s Astrophysics Data System.

\end{acknowledgements}

\bibliographystyle{aa}
\bibliography{bibliography}

\begin{thebibliography}{48}
\expandafter\ifx\csname natexlab\endcsname\relax\def\natexlab#1{#1}\fi

\bibitem[{{Airapetian} {et~al.}(2010){Airapetian}, {Carpenter}, \&
  {Ofman}}]{2010ApJ...723.1210A}
{Airapetian}, V., {Carpenter}, K.~G., \& {Ofman}, L. 2010, \apj, 723, 1210

\bibitem[{{Alvarez} {et~al.}(2001){Alvarez}, {Jorissen}, {Plez}, {Gillet},
  {Fokin}, \& {Dedecker}}]{2001A&A...379..288A}
{Alvarez}, R., {Jorissen}, A., {Plez}, B., {et~al.} 2001, \aap, 379, 288

\bibitem[{{Arroyo-Torres} {et~al.}(2015){Arroyo-Torres}, {Wittkowski},
  {Chiavassa}, {Scholz}, {Freytag}, {Marcaide}, {Hauschildt}, {Wood}, \&
  {Abellan}}]{2015A&A...575A..50A}
{Arroyo-Torres}, B., {Wittkowski}, M., {Chiavassa}, A., {et~al.} 2015, \aap,
  575, A50

\bibitem[{{Asplund} {et~al.}(2009){Asplund}, {Grevesse}, {Sauval}, \&
  {Scott}}]{2009ARA&A..47..481A}
{Asplund}, M., {Grevesse}, N., {Sauval}, A.~J., \& {Scott}, P. 2009, \araa, 47,
  481

\bibitem[{{Boboltz} \& {Wittkowski}(2005)}]{2005ApJ...618..953B}
{Boboltz}, D.~A. \& {Wittkowski}, M. 2005, \apj, 618, 953

\bibitem[{{Chelli} {et~al.}(2016){Chelli}, {Duvert}, {Bourg{\`e}s}, {Mella},
  {Lafrasse}, {Bonneau}, \& {Chesneau}}]{2016A&A...589A.112C}
{Chelli}, A., {Duvert}, G., {Bourg{\`e}s}, L., {et~al.} 2016, \aap, 589, A112

\bibitem[{{Chelli} {et~al.}(2009){Chelli}, {Utrera}, \&
  {Duvert}}]{2009A&A...502..705C}
{Chelli}, A., {Utrera}, O.~H., \& {Duvert}, G. 2009, \aap, 502, 705

\bibitem[{{Chiavassa} {et~al.}(2018){Chiavassa}, {Freytag}, \&
  {Schultheis}}]{2018A&A...617L...1C}
{Chiavassa}, A., {Freytag}, B., \& {Schultheis}, M. 2018, \aap, 617, L1

\bibitem[{{Chiavassa} {et~al.}(2010){Chiavassa}, {Lacour}, {Millour}, {Driebe},
  {Wittkowski}, {Plez}, {Thi{\'e}baut}, {Josselin}, {Freytag}, {Scholz}, \&
  {Haubois}}]{2010A&A...511A..51C}
{Chiavassa}, A., {Lacour}, S., {Millour}, F., {et~al.} 2010, \aap, 511, A51

\bibitem[{{Chiavassa} {et~al.}(2009){Chiavassa}, {Plez}, {Josselin}, \&
  {Freytag}}]{2009A&A...506.1351C}
{Chiavassa}, A., {Plez}, B., {Josselin}, E., \& {Freytag}, B. 2009, \aap, 506,
  1351

\bibitem[{{Driebe} {et~al.}(2009){Driebe}, {Groh}, {Hofmann}, {Ohnaka},
  {Kraus}, {Millour}, {Murakawa}, {Schertl}, {Weigelt}, {Petrov}, {Wittkowski},
  {Hummel}, {Le Bouquin}, {Merand}, {Sch{\"o}ller}, {Massi}, {Stee}, \&
  {Tatulli}}]{2009A&A...507..301D}
{Driebe}, T., {Groh}, J.~H., {Hofmann}, K.~H., {et~al.} 2009, \aap, 507, 301

\bibitem[{{Follert} {et~al.}(2014){Follert}, {Dorn}, {Oliva}, {Lizon},
  {Hatzes}, {Piskunov}, {Reiners}, {Seemann}, {Stempels}, {Heiter}, {Marquart},
  {Lockhart}, {Anglada-Escude}, {L{\"o}winger}, {Baade}, {Grunhut}, {Bristow},
  {Klein}, {Jung}, {Ives}, {Kerber}, {Pozna}, {Paufique}, {Kaeufl}, {Origlia},
  {Valenti}, {Gojak}, {Hilker}, {Pasquini}, {Smette}, \&
  {Smoker}}]{2014SPIE.9147E..19F}
{Follert}, R., {Dorn}, R.~J., {Oliva}, E., {et~al.} 2014, Society of
  Photo-Optical Instrumentation Engineers (SPIE) Conference Series, Vol. 9147,
  {CRIRES+: a cross-dispersed high-resolution infrared spectrograph for the ESO
  VLT}, 914719

\bibitem[{{Freytag} {et~al.}(2017){Freytag}, {Liljegren}, \&
  {H{\"o}fner}}]{2017A&A...600A.137F}
{Freytag}, B., {Liljegren}, S., \& {H{\"o}fner}, S. 2017, \aap, 600, A137

\bibitem[{{Freytag} {et~al.}(2012){Freytag}, {Steffen}, {Ludwig},
  {Wedemeyer-B{\"o}hm}, {Schaffenberger}, \& {Steiner}}]{2012JCoPh.231..919F}
{Freytag}, B., {Steffen}, M., {Ludwig}, H.-G., {et~al.} 2012, Journal of
  Computational Physics, 231, 919

\bibitem[{{Gaia Collaboration} {et~al.}(2018){Gaia Collaboration}, {Brown},
  {Vallenari}, {Prusti}, {de Bruijne}, {Babusiaux}, {Bailer-Jones}, {Biermann},
  {Evans}, {Eyer}, \& et~al.}]{2018A&A...616A...1G}
{Gaia Collaboration}, {Brown}, A.~G.~A., {Vallenari}, A., {et~al.} 2018, \aap,
  616, A1

\bibitem[{{Gaia Collaboration} {et~al.}(2016){Gaia Collaboration}, {Prusti},
  {de Bruijne}, {Brown}, {Vallenari}, {Babusiaux}, {Bailer-Jones}, {Bastian},
  {Biermann}, {Evans}, \& et~al.}]{2016A&A...595A...1G}
{Gaia Collaboration}, {Prusti}, T., {de Bruijne}, J.~H.~J., {et~al.} 2016,
  \aap, 595, A1

\bibitem[{{Gravity Collaboration} {et~al.}(2017){Gravity Collaboration},
  {Abuter}, {Accardo}, {Amorim}, {Anugu}, {{\'A}vila}, {Azouaoui}, {Benisty},
  {Berger}, {Blind}, {Bonnet}, {Bourget}, {Brandner}, {Brast}, {Buron},
  {Burtscher}, {Cassaing}, {Chapron}, {Choquet}, {Cl{\'e}net}, {Collin},
  {Coud{\'e} Du Foresto}, {de Wit}, {de Zeeuw}, {Deen},
  {Delplancke-Str{\"o}bele}, {Dembet}, {Derie}, {Dexter}, {Duvert}, {Ebert},
  {Eckart}, {Eisenhauer}, {Esselborn}, {F{\'e}dou}, {Finger}, {Garcia}, {Garcia
  Dabo}, {Garcia Lopez}, {Gendron}, {Genzel}, {Gillessen}, {Gonte}, {Gordo},
  {Grould}, {Gr{\"o}zinger}, {Guieu}, {Haguenauer}, {Hans}, {Haubois}, {Haug},
  {Haussmann}, {Henning}, {Hippler}, {Horrobin}, {Huber}, {Hubert}, {Hubin},
  {Hummel}, {Jakob}, {Janssen}, {Jochum}, {Jocou}, {Kaufer}, {Kellner},
  {Kendrew}, {Kern}, {Kervella}, {Kiekebusch}, {Klein}, {Kok}, {Kolb}, {Kulas},
  {Lacour}, {Lapeyr{\`e}re}, {Lazareff}, {Le Bouquin}, {L{\`e}na}, {Lenzen},
  {L{\'e}v{\^e}que}, {Lippa}, {Magnard}, {Mehrgan}, {Mellein}, {M{\'e}rand},
  {Moreno-Ventas}, {Moulin}, {M{\"u}ller}, {M{\"u}ller}, {Neumann}, {Oberti},
  {Ott}, {Pallanca}, {Panduro}, {Pasquini}, {Paumard}, {Percheron}, {Perraut},
  {Perrin}, {Pfl{\"u}ger}, {Pfuhl}, {Phan Duc}, {Plewa}, {Popovic}, {Rabien},
  {Ram{\'{\i}}rez}, {Ramos}, {Rau}, {Riquelme}, {Rohloff}, {Rousset},
  {Sanchez-Bermudez}, {Scheithauer}, {Sch{\"o}ller}, {Schuhler}, {Spyromilio},
  {Straubmeier}, {Sturm}, {Suarez}, {Tristram}, {Ventura}, {Vincent},
  {Waisberg}, {Wank}, {Weber}, {Wieprecht}, {Wiest}, {Wiezorrek}, {Wittkowski},
  {Woillez}, {Wolff}, {Yazici}, {Ziegler}, \& {Zins}}]{2017A&A...602A..94G}
{Gravity Collaboration}, {Abuter}, R., {Accardo}, M., {et~al.} 2017, \aap, 602,
  A94

\bibitem[{{Gustafsson} {et~al.}(2008){Gustafsson}, {Edvardsson}, {Eriksson},
  {J{\o}rgensen}, {Nordlund}, \& {Plez}}]{2008A&A...486..951G}
{Gustafsson}, B., {Edvardsson}, B., {Eriksson}, K., {et~al.} 2008, \aap, 486,
  951

\bibitem[{{H{\"o}fner} \& {Freytag}(2019)}]{2019A&A...623A.158H}
{H{\"o}fner}, S. \& {Freytag}, B. 2019, \aap, 623, A158

\bibitem[{{H{\"o}fner} \& {Olofsson}(2018)}]{2018A&ARv..26....1H}
{H{\"o}fner}, S. \& {Olofsson}, H. 2018, \aapr, 26, 1

\bibitem[{{Ireland} {et~al.}(2008){Ireland}, {Scholz}, \&
  {Wood}}]{2008MNRAS.391.1994I}
{Ireland}, M.~J., {Scholz}, M., \& {Wood}, P.~R. 2008, \mnras, 391, 1994

\bibitem[{{Ireland} {et~al.}(2011){Ireland}, {Scholz}, \&
  {Wood}}]{2011MNRAS.418..114I}
{Ireland}, M.~J., {Scholz}, M., \& {Wood}, P.~R. 2011, \mnras, 418, 114

\bibitem[{{Jones} {et~al.}(2013){Jones}, {Noll}, {Kausch}, {Szyszka}, \&
  {Kimeswenger}}]{2013A&A...560A..91J}
{Jones}, A., {Noll}, S., {Kausch}, W., {Szyszka}, C., \& {Kimeswenger}, S.
  2013, \aap, 560, A91

\bibitem[{{Kafka}(2018)}]{AAVSO}
{Kafka}, S. 2018, Observations from the AAVSO International Database,
  https://www.aavso.org

\bibitem[{{Kravchenko} {et~al.}(2019){Kravchenko}, {Chiavassa}, {Van Eck},
  {Jorissen}, {Merle}, {Freytag}, \& {Plez}}]{2019A&A...632A..28K}
{Kravchenko}, K., {Chiavassa}, A., {Van Eck}, S., {et~al.} 2019, \aap, 632, A28

\bibitem[{{Kravchenko} {et~al.}(2018){Kravchenko}, {Van Eck}, {Chiavassa},
  {Jorissen}, {Freytag}, \& {Plez}}]{2018A&A...610A..29K}
{Kravchenko}, K., {Van Eck}, S., {Chiavassa}, A., {et~al.} 2018, \aap, 610, A29

\bibitem[{{Le Bouquin} {et~al.}(2008){Le Bouquin}, {Abuter}, {Bauvir},
  {Bonnet}, {Haguenauer}, {di Lieto}, {Menardi}, {Morel}, {Rantakyr{\"o}},
  {Schoeller}, {Wallander}, \& {Wehner}}]{2008SPIE.7013E..18L}
{Le Bouquin}, J.-B., {Abuter}, R., {Bauvir}, B., {et~al.} 2008, in \procspie,
  Vol. 7013, Optical and Infrared Interferometry, 701318

\bibitem[{{Liljegren} {et~al.}(2016){Liljegren}, {H{\"o}fner}, {Nowotny}, \&
  {Eriksson}}]{2016A&A...589A.130L}
{Liljegren}, S., {H{\"o}fner}, S., {Nowotny}, W., \& {Eriksson}, K. 2016, \aap,
  589, A130

\bibitem[{{Magain}(1986)}]{1986A&A...163..135M}
{Magain}, P. 1986, \aap, 163, 135

\bibitem[{{Millan-Gabet} {et~al.}(2005){Millan-Gabet}, {Pedretti}, {Monnier},
  {Schloerb}, {Traub}, {Carleton}, {Lacasse}, \&
  {Segransan}}]{2005ApJ...620..961M}
{Millan-Gabet}, R., {Pedretti}, E., {Monnier}, J.~D., {et~al.} 2005, \apj, 620,
  961

\bibitem[{{Noll} {et~al.}(2012){Noll}, {Kausch}, {Barden}, {Jones}, {Szyszka},
  {Kimeswenger}, \& {Vinther}}]{2012A&A...543A..92N}
{Noll}, S., {Kausch}, W., {Barden}, M., {et~al.} 2012, \aap, 543, A92

\bibitem[{{Ohnaka} {et~al.}(2019){Ohnaka}, {Hadjara}, \& {Maluenda
  Berna}}]{2019A&A...621A...6O}
{Ohnaka}, K., {Hadjara}, M., \& {Maluenda Berna}, M.~Y.~L. 2019, \aap, 621, A6

\bibitem[{{Ohnaka} {et~al.}(2009){Ohnaka}, {Hofmann}, {Benisty}, {Chelli},
  {Driebe}, {Millour}, {Petrov}, {Schertl}, {Stee}, {Vakili}, \&
  {Weigelt}}]{2009A&A...503..183O}
{Ohnaka}, K., {Hofmann}, K.-H., {Benisty}, M., {et~al.} 2009, \aap, 503, 183

\bibitem[{{Ohnaka} {et~al.}(2016){Ohnaka}, {Weigelt}, \&
  {Hofmann}}]{2016A&A...589A..91O}
{Ohnaka}, K., {Weigelt}, G., \& {Hofmann}, K.-H. 2016, \aap, 589, A91

\bibitem[{{Petrov} {et~al.}(2007){Petrov}, {Malbet}, {Weigelt}, {Antonelli},
  {Beckmann}, {Bresson}, {Chelli}, {Dugu{\'e}}, {Duvert}, {Gennari},
  {Gl{\"u}ck}, {Kern}, {Lagarde}, {Le Coarer}, {Lisi}, {Millour}, {Perraut},
  {Puget}, {Rantakyr{\"o}}, {Robbe-Dubois}, {Roussel}, {Salinari}, {Tatulli},
  {Zins}, {Accardo}, {Acke}, {Agabi}, {Altariba}, {Arezki}, {Aristidi},
  {Baffa}, {Behrend}, {Bl{\"o}cker}, {Bonhomme}, {Busoni}, {Cassaing},
  {Clausse}, {Colin}, {Connot}, {Delboulb{\'e}}, {Domiciano de Souza},
  {Driebe}, {Feautrier}, {Ferruzzi}, {Forveille}, {Fossat}, {Foy},
  {Fraix-Burnet}, {Gallardo}, {Giani}, {Gil}, {Glentzlin}, {Heiden},
  {Heininger}, {Hernandez Utrera}, {Hofmann}, {Kamm}, {Kiekebusch}, {Kraus},
  {Le Contel}, {Le Contel}, {Lesourd}, {Lopez}, {Lopez}, {Magnard}, {Marconi},
  {Mars}, {Martinot-Lagarde}, {Mathias}, {M{\`e}ge}, {Monin}, {Mouillet},
  {Mourard}, {Nussbaum}, {Ohnaka}, {Pacheco}, {Perrier}, {Rabbia}, {Rebattu},
  {Reynaud}, {Richichi}, {Robini}, {Sacchettini}, {Schertl}, {Sch{\"o}ller},
  {Solscheid}, {Spang}, {Stee}, {Stefanini}, {Tallon}, {Tallon-Bosc}, {Tasso},
  {Testi}, {Vakili}, {von der L{\"u}he}, {Valtier}, {Vannier}, \&
  {Ventura}}]{2007A&A...464....1P}
{Petrov}, R.~G., {Malbet}, F., {Weigelt}, G., {et~al.} 2007, \aap, 464, 1

\bibitem[{{Samus'} {et~al.}(2017){Samus'}, {Kazarovets}, {Durlevich},
  {Kireeva}, \& {Pastukhova}}]{2017ARep...61...80S}
{Samus'}, N.~N., {Kazarovets}, E.~V., {Durlevich}, O.~V., {Kireeva}, N.~N., \&
  {Pastukhova}, E.~N. 2017, Astronomy Reports, 61, 80

\bibitem[{{Tatulli} {et~al.}(2007){Tatulli}, {Millour}, {Chelli}, {Duvert},
  {Acke}, {Hernandez Utrera}, {Hofmann}, {Kraus}, {Malbet}, {M{\`e}ge},
  {Petrov}, {Vannier}, {Zins}, {Antonelli}, {Beckmann}, {Bresson}, {Dugu{\'e}},
  {Gennari}, {Gl{\"u}ck}, {Kern}, {Lagarde}, {Le Coarer}, {Lisi}, {Perraut},
  {Puget}, {Rantakyr{\"o}}, {Robbe-Dubois}, {Roussel}, {Weigelt}, {Accardo},
  {Agabi}, {Altariba}, {Arezki}, {Aristidi}, {Baffa}, {Behrend}, {Bl{\"o}cker},
  {Bonhomme}, {Busoni}, {Cassaing}, {Clausse}, {Colin}, {Connot},
  {Delboulb{\'e}}, {Domiciano de Souza}, {Driebe}, {Feautrier}, {Ferruzzi},
  {Forveille}, {Fossat}, {Foy}, {Fraix-Burnet}, {Gallardo}, {Giani}, {Gil},
  {Glentzlin}, {Heiden}, {Heininger}, {Kamm}, {Kiekebusch}, {Le Contel}, {Le
  Contel}, {Lesourd}, {Lopez}, {Lopez}, {Magnard}, {Marconi}, {Mars},
  {Martinot-Lagarde}, {Mathias}, {Monin}, {Mouillet}, {Mourard}, {Nussbaum},
  {Ohnaka}, {Pacheco}, {Perrier}, {Rabbia}, {Rebattu}, {Reynaud}, {Richichi},
  {Robini}, {Sacchettini}, {Schertl}, {Sch{\"o}ller}, {Solscheid}, {Spang},
  {Stee}, {Stefanini}, {Tallon}, {Tallon-Bosc}, {Tasso}, {Testi}, {Vakili},
  {von der L{\"u}he}, {Valtier}, \& {Ventura}}]{2007A&A...464...29T}
{Tatulli}, E., {Millour}, F., {Chelli}, A., {et~al.} 2007, \aap, 464, 29

\bibitem[{{Templeton} {et~al.}(2005){Templeton}, {Mattei}, \&
  {Willson}}]{2005AJ....130..776T}
{Templeton}, M.~R., {Mattei}, J.~A., \& {Willson}, L.~A. 2005, \aj, 130, 776

\bibitem[{{Tsuji}(2000)}]{2000ApJ...540L..99T}
{Tsuji}, T. 2000, \apjl, 540, L99

\bibitem[{{Uttenthaler} {et~al.}(2011){Uttenthaler}, {van Stiphout}, {Voet},
  {van Winckel}, {van Eck}, {Jorissen}, {Kerschbaum}, {Raskin}, {Prins},
  {Pessemier}, {Waelkens}, {Fr{\'e}mat}, {Hensberge}, {Dumortier}, \&
  {Lehmann}}]{2011A&A...531A..88U}
{Uttenthaler}, S., {van Stiphout}, K., {Voet}, K., {et~al.} 2011, \aap, 531,
  A88

\bibitem[{{van Belle} {et~al.}(1996){van Belle}, {Dyck}, {Benson}, \&
  {Lacasse}}]{1996AJ....112.2147V}
{van Belle}, G.~T., {Dyck}, H.~M., {Benson}, J.~A., \& {Lacasse}, M.~G. 1996,
  \aj, 112, 2147

\bibitem[{{Vigan} {et~al.}(2018){Vigan}, {Otten}, {Muslimov}, {Dohlen},
  {Philipps}, {Seemann}, {Beuzit}, {Dorn}, {Kasper}, {Mouillet}, {Baraffe}, \&
  {Reiners}}]{2018SPIE10702E..36V}
{Vigan}, A., {Otten}, G.~P.~P.~L., {Muslimov}, E., {et~al.} 2018, in Society of
  Photo-Optical Instrumentation Engineers (SPIE) Conference Series, Vol. 10702,
  \procspie, 1070236

\bibitem[{{Whitelock} {et~al.}(2008){Whitelock}, {Feast}, \& {Van
  Leeuwen}}]{2008MNRAS.386..313W}
{Whitelock}, P.~A., {Feast}, M.~W., \& {Van Leeuwen}, F. 2008, \mnras, 386, 313

\bibitem[{{Wittkowski} {et~al.}(2017){Wittkowski}, {Arroyo-Torres}, {Marcaide},
  {Abellan}, {Chiavassa}, \& {Guirado}}]{2017A&A...597A...9W}
{Wittkowski}, M., {Arroyo-Torres}, B., {Marcaide}, J.~M., {et~al.} 2017, \aap,
  597, A9

\bibitem[{{Wittkowski} {et~al.}(2008){Wittkowski}, {Boboltz}, {Driebe}, {Le
  Bouquin}, {Millour}, {Ohnaka}, \& {Scholz}}]{2008A&A...479L..21W}
{Wittkowski}, M., {Boboltz}, D.~A., {Driebe}, T., {et~al.} 2008, \aap, 479, L21

\bibitem[{{Wittkowski} {et~al.}(2011){Wittkowski}, {Boboltz}, {Ireland},
  {Karovicova}, {Ohnaka}, {Scholz}, {van Wyk}, {Whitelock}, {Wood}, \&
  {Zijlstra}}]{2011A&A...532L...7W}
{Wittkowski}, M., {Boboltz}, D.~A., {Ireland}, M., {et~al.} 2011, \aap, 532, L7

\bibitem[{{Wittkowski} {et~al.}(2016){Wittkowski}, {Chiavassa}, {Freytag},
  {Scholz}, {H{\"o}fner}, {Karovicova}, \& {Whitelock}}]{2016A&A...587A..12W}
{Wittkowski}, M., {Chiavassa}, A., {Freytag}, B., {et~al.} 2016, \aap, 587, A12

\bibitem[{{Wittkowski} {et~al.}(2018){Wittkowski}, {Rau}, {Chiavassa},
  {H{\"o}fner}, {Scholz}, {Wood}, {de Wit}, {Eisenhauer}, {Haubois}, \&
  {Paumard}}]{2018A&A...613L...7W}
{Wittkowski}, M., {Rau}, G., {Chiavassa}, A., {et~al.} 2018, \aap, 613, L7

\end{thebibliography}

\begin{appendix}

\section{Additional figures}
\label{app:figures}

\begin{figure}[h]
   \centering
   \includegraphics[width=8.5cm]{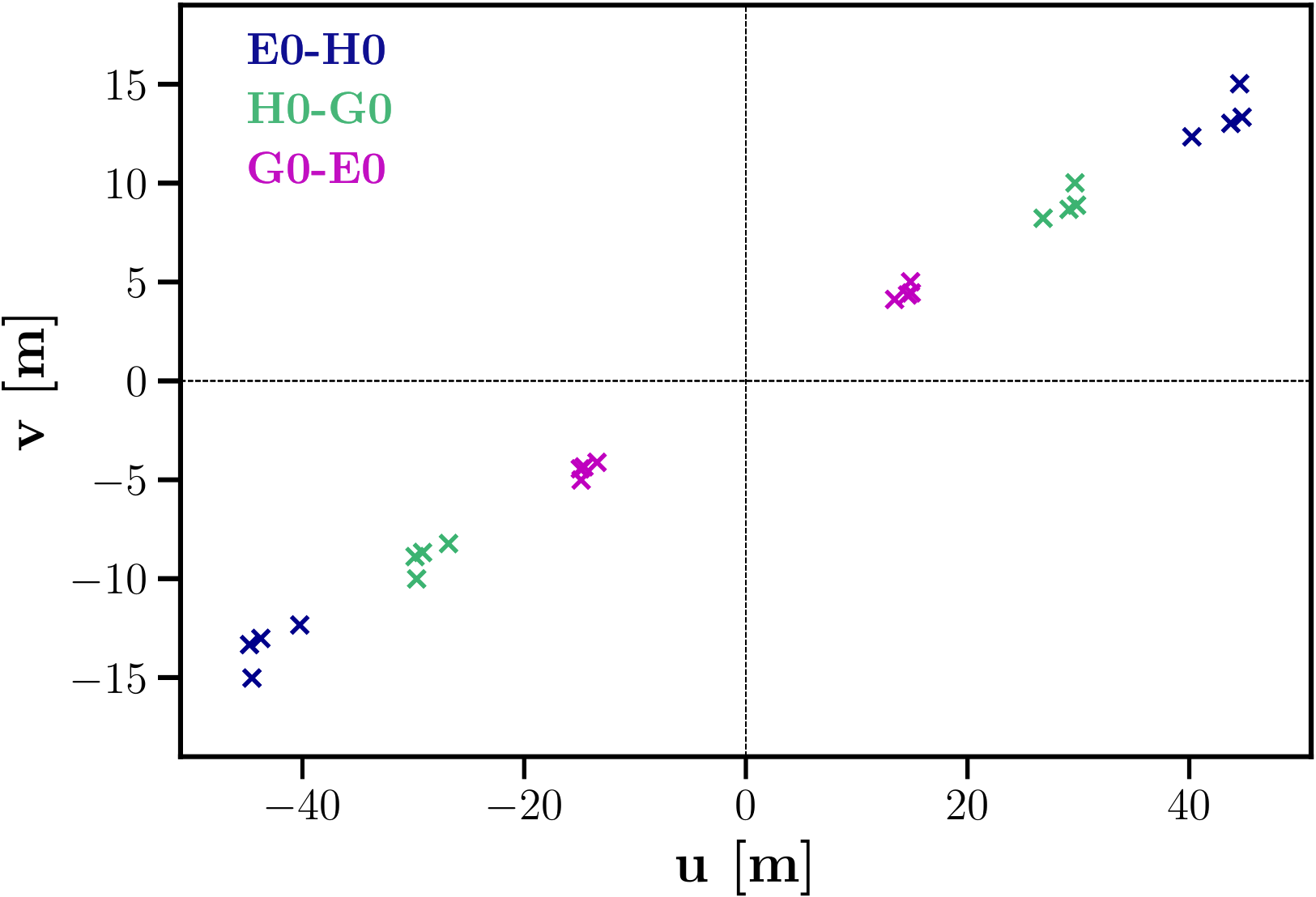}\\
   \includegraphics[width=8.5cm]{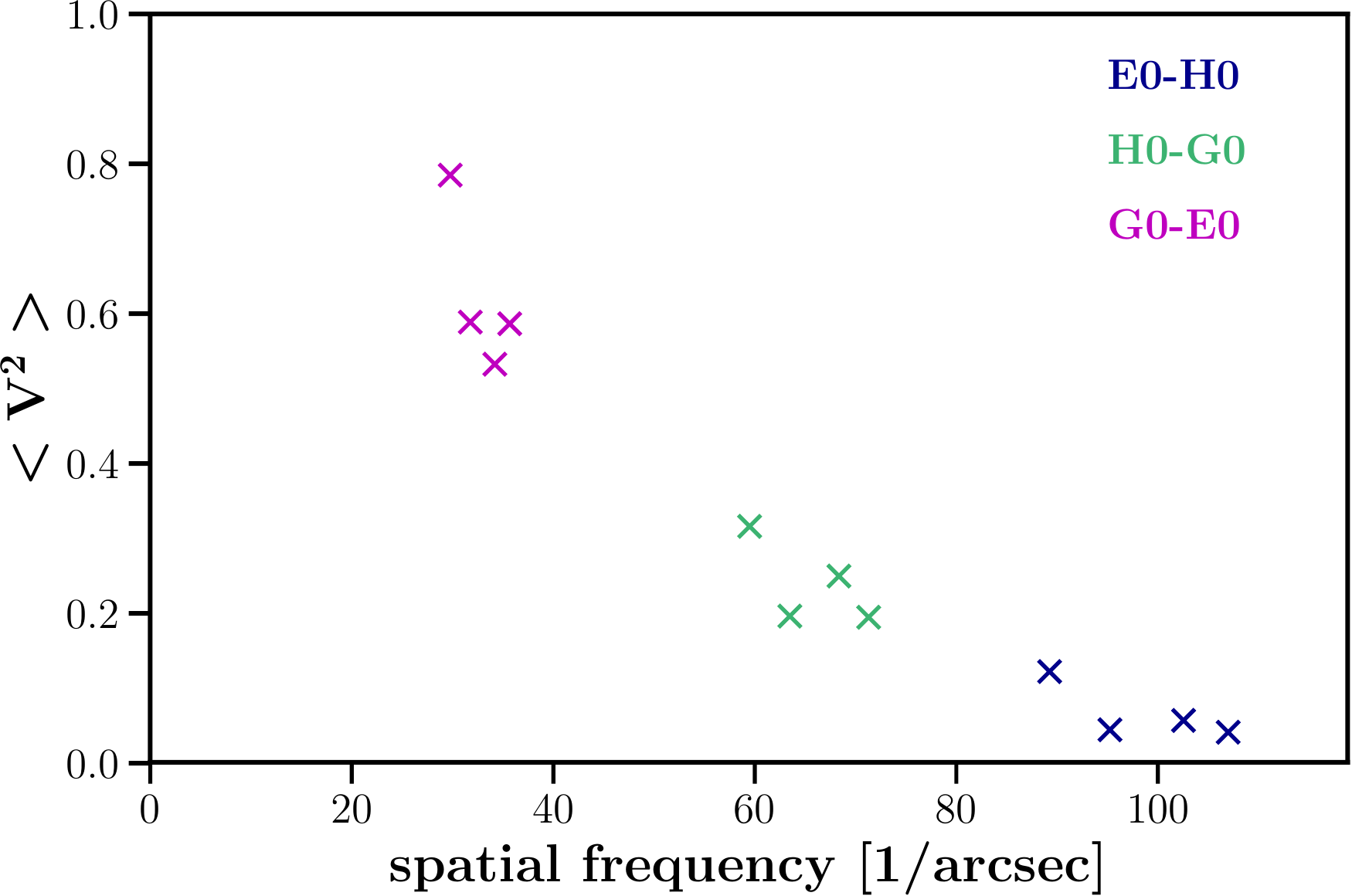}
   \caption{\textit{Top panel:} $uv$-plane of AMBER observations. Colors correspond to different baselines. \textit{Bottom panel:} Average squared visibility amplitudes $<V^2>$ versus spatial frequencies for the three baselines and different observing dates.} 
            \label{Fig:uv_plot}
   \end{figure}

\begin{figure*}[h]
   \centering
   \includegraphics[width=12cm]{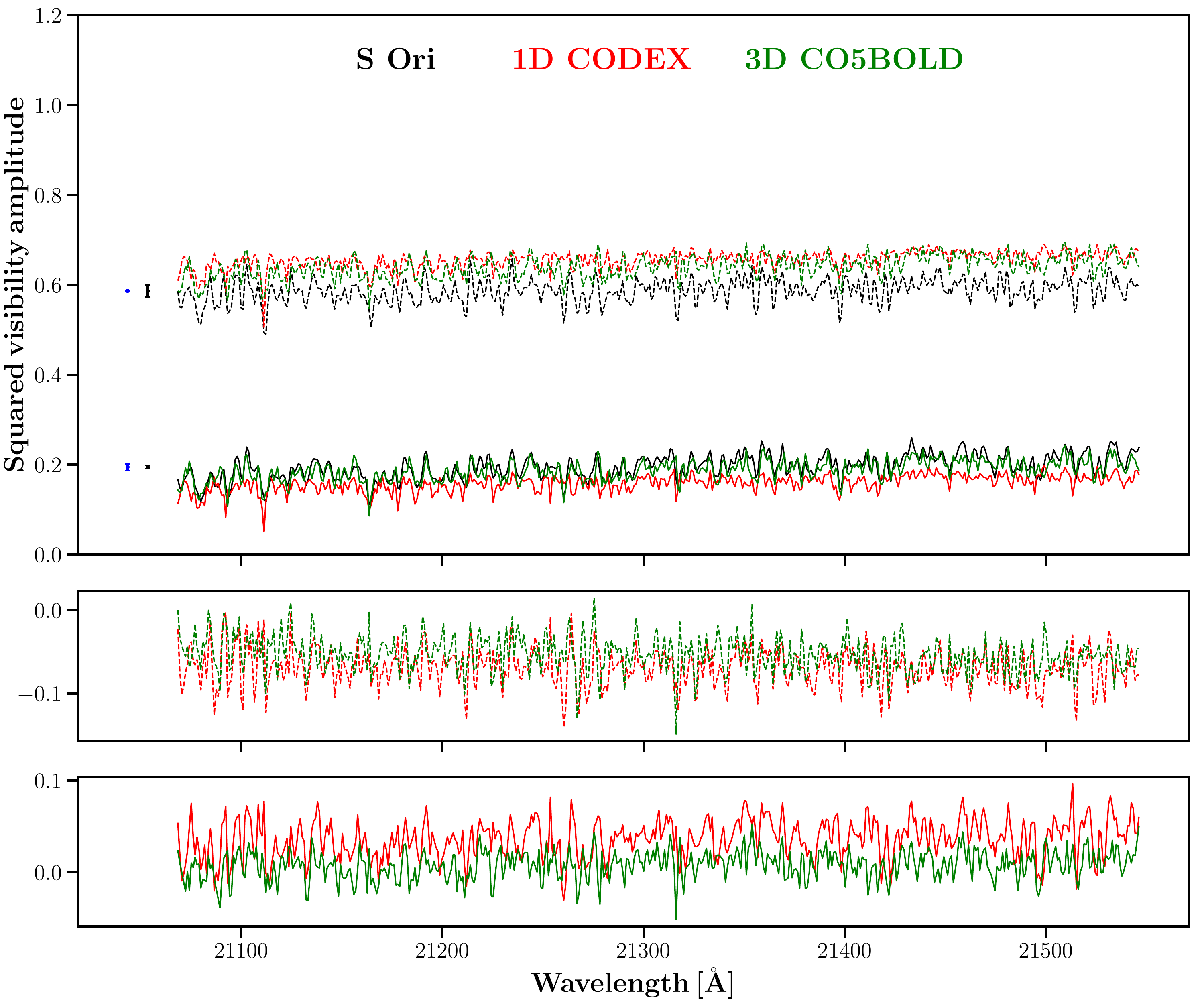}    %17cm
   \caption{\textit{Top panel:} Squared visibility amplitudes of S Ori (black) obtained with the 2.133 $\mu$m AMBER setting (Table~\ref{tab:data_overview}). They are compared to the best-fit 1D CODEX snapshot (red) and the best-fit snapshot from the 3D RHD CO5BOLD simulation (green) described in Sects.~\ref{CODEX_UD} and \ref{3D_UD} respectively. Only data corresponding to the short (G0-E0, dashed lines) and medium (H0-G0, solid lines) AMBER baselines are shown. Average errorbars (mainly arising from the uncertainty on the angular diameter of the calibrator) of AMBER observations are shown  in black on the left side of each visibility spectrum. Systematic errorbars associated with the percentage of the best frames being averaged (see Sect.~\ref{Sect:observations}) are shown in blue. Errorbars which are too small to discern are shown as ten-fold errors and labeled accordingly. \textit{Middle panel:} Residuals between the AMBER and model squared visibilities for the short baseline. The color coding is the same as in the top panel. \textit{Bottom panel:} Same as the middle panel, but for the medium baseline.}
            \label{Fig:data_vs_UD_1}
   \end{figure*}

\begin{figure*}[h]
   \centering
   \includegraphics[width=12cm]{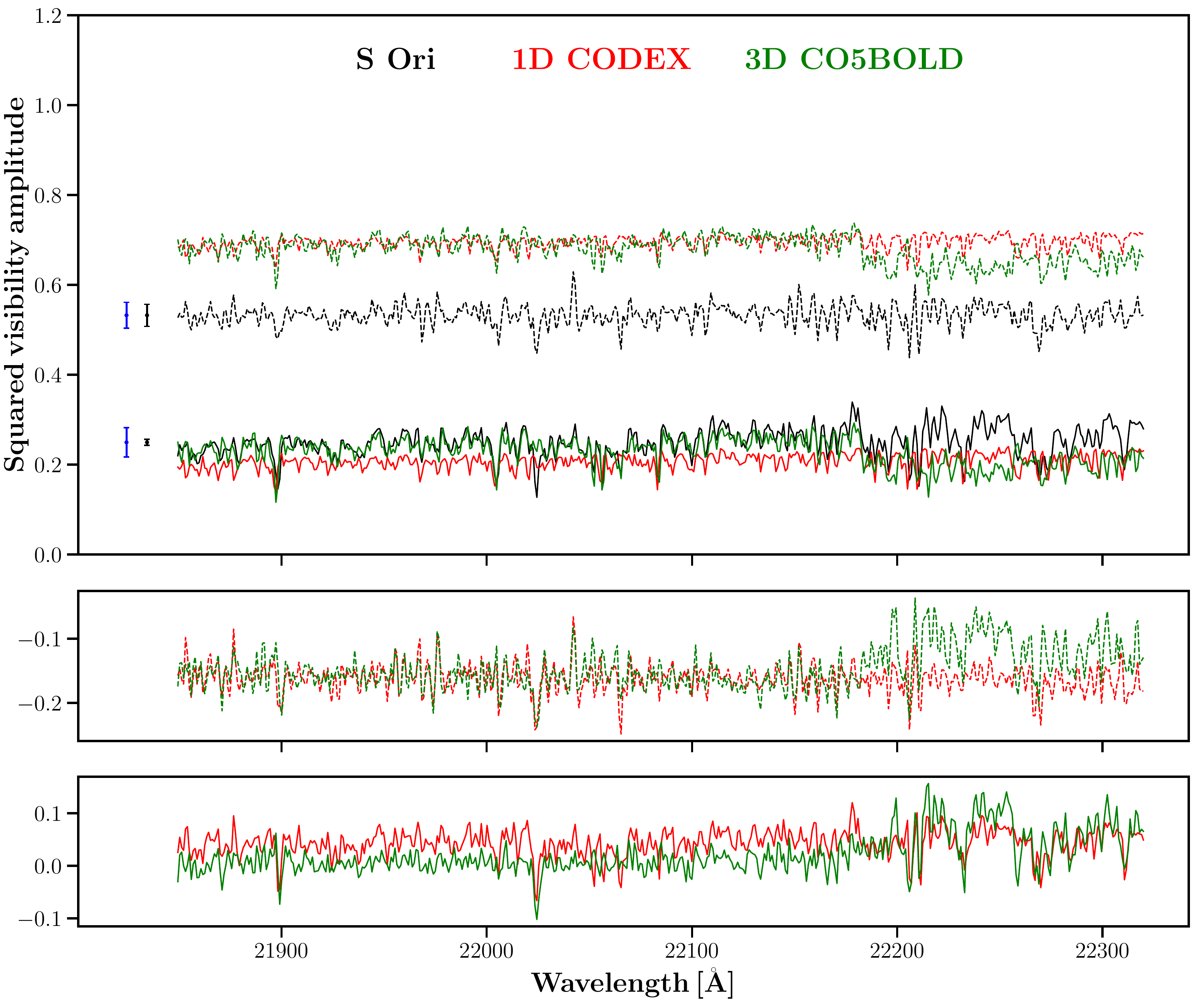}
   \caption{Same as Fig.~\ref{Fig:data_vs_UD_1}, but for data obtained with the 2.211 $\mu$m AMBER setting (Table~\ref{tab:data_overview}).}
            \label{Fig:data_vs_UD_3}
   \end{figure*}

\begin{figure*}[h]
   \centering
   \includegraphics[width=12cm]{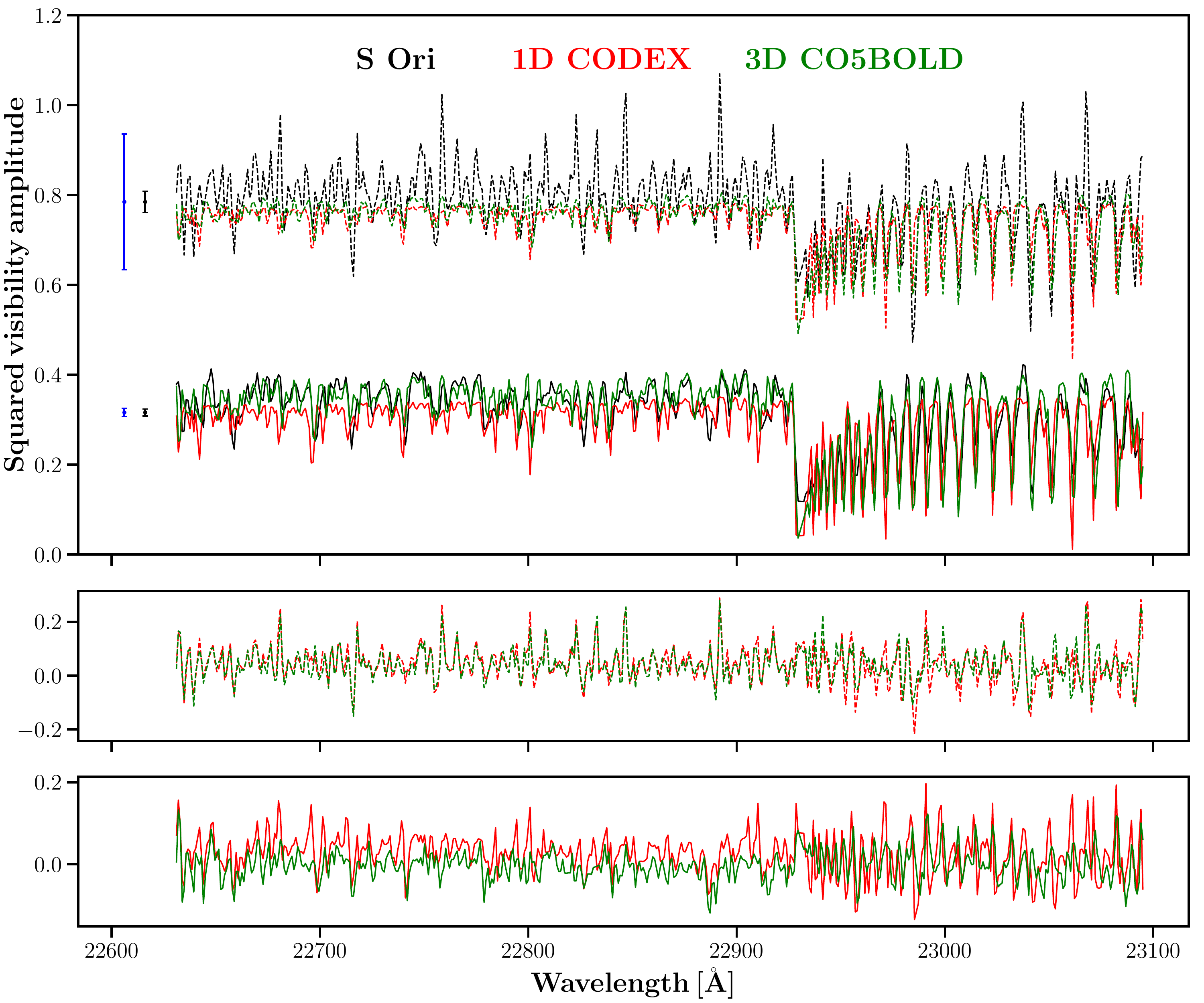}
   \caption{Same as Fig.~\ref{Fig:data_vs_UD_1}, but for data obtained with the 2.288 $\mu$m AMBER setting (Table~\ref{tab:data_overview}).}
            \label{Fig:data_vs_UD_4}
   \end{figure*}

\begin{figure*}[h]
   \centering
   \includegraphics[width=12cm]{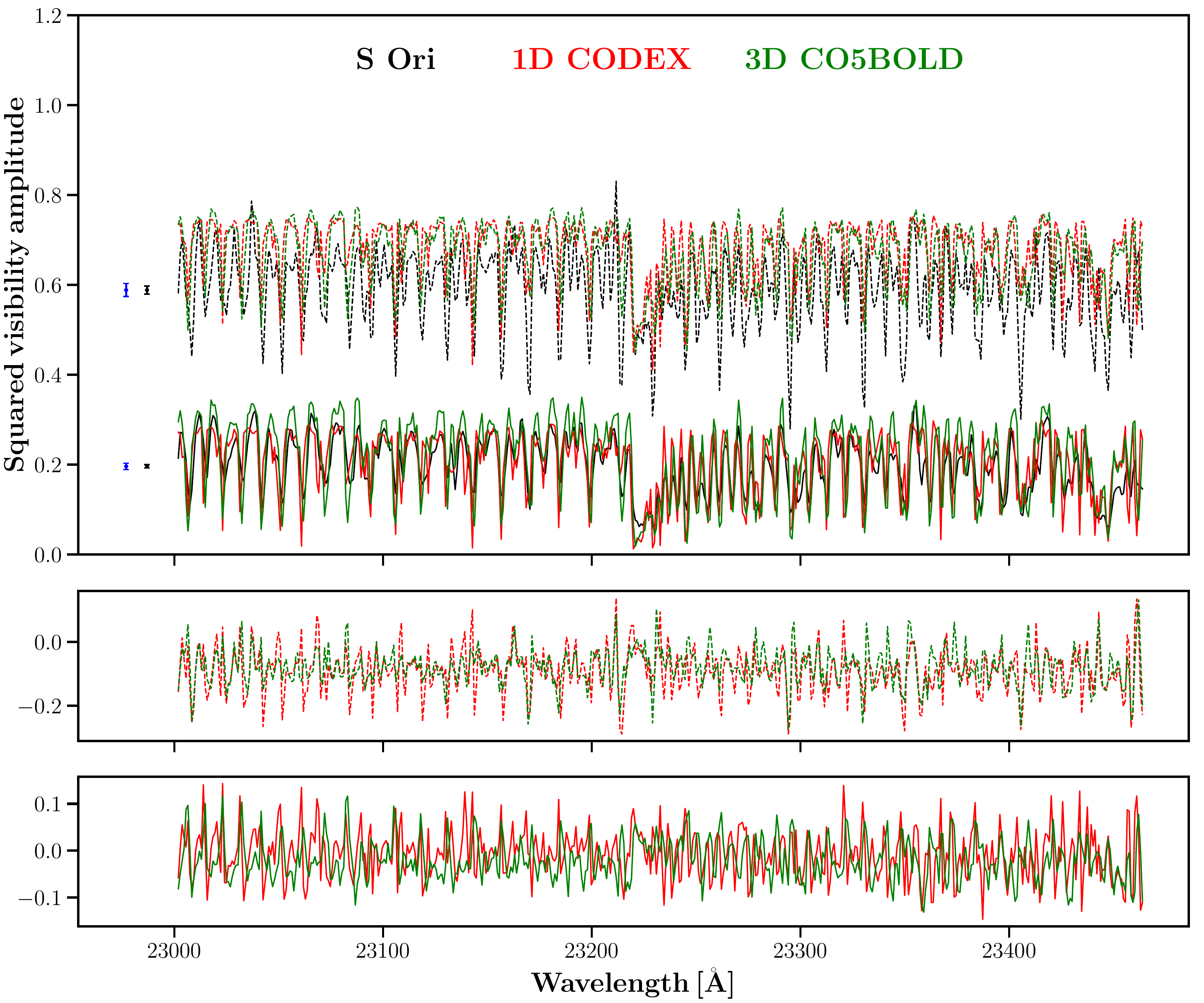}
   \caption{Same as Fig.~\ref{Fig:data_vs_UD_1}, but for data obtained with the 2.326 $\mu$m AMBER setting (Table~\ref{tab:data_overview}).}
            \label{Fig:data_vs_UD_5}
   \end{figure*}

\begin{figure}[h]
   \centering
   \includegraphics[width=9cm]{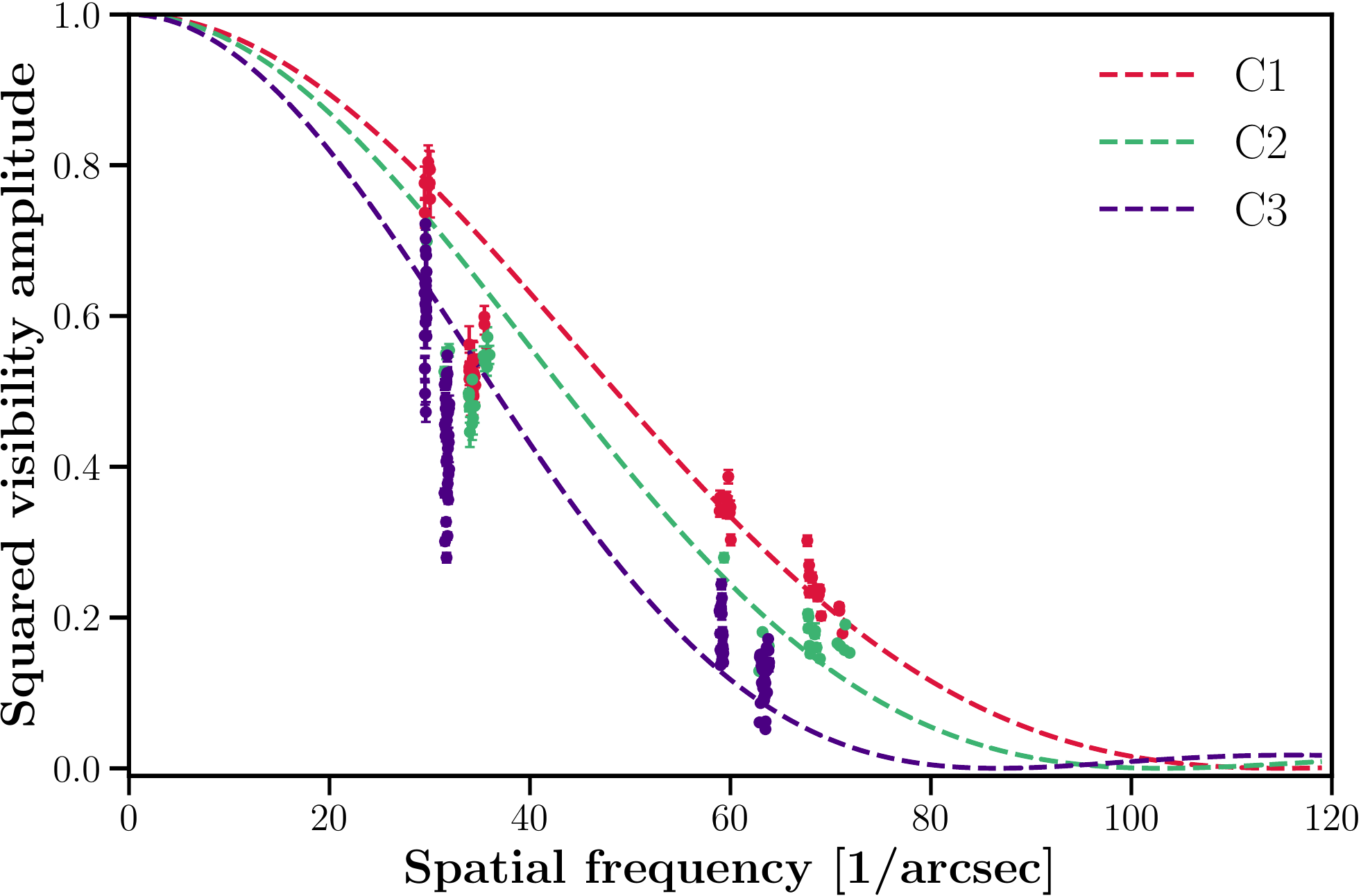}
   \caption{UD fit of VLTI/AMBER visibilities of S~Ori (dashed lines) for the different tomographic masks. Two clusters of points correspond to two baselines: small (G0-E0) and medium (H0-G0). Each point corresponds to a squared visibility extracted at a wavelength contributing to a given tomographic mask. Colors correspond to different masks.} 
            \label{Fig:AMBER_spatfreq_plot}
   \end{figure}

\begin{figure}[h]
   \centering
   \includegraphics[width=9cm]{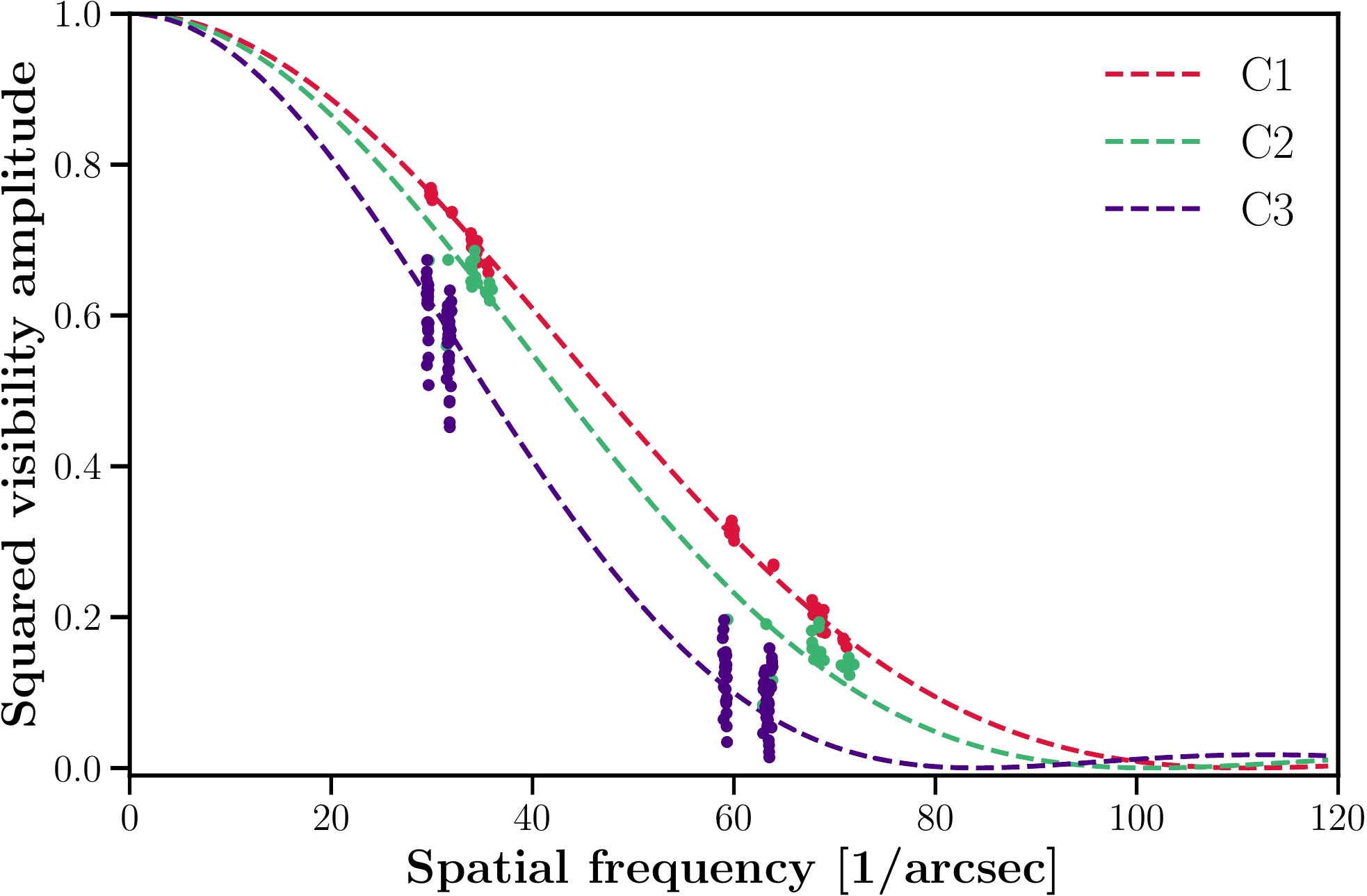}
   \caption{Same as Fig.~\ref{Fig:AMBER_spatfreq_plot}, but for the best-matching CODEX model (Sect.~\ref{CODEX_UD}).}
            \label{Fig:CODEX_spatfreq_plot}
   \end{figure}

\begin{figure}[h]
   \centering
   \includegraphics[width=9cm]{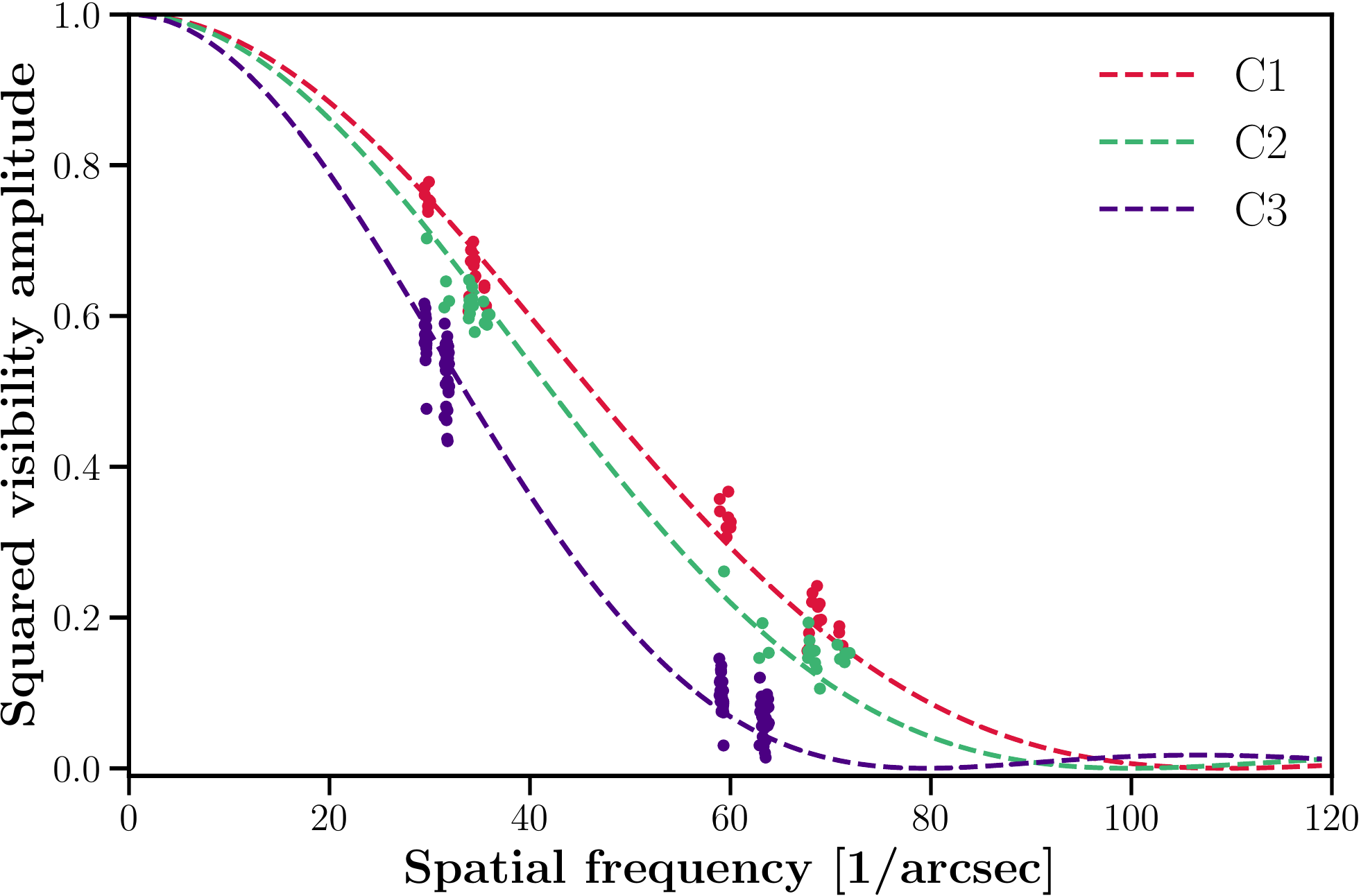}
   \caption{Same as Fig.~\ref{Fig:AMBER_spatfreq_plot}, but for the best-matching 3D snapshot (Sect.~\ref{3D_UD}).} 
            \label{Fig:CO5BOLD_spatfreq_plot}
   \end{figure}

Here, we provide additional figures. Figure~\ref{Fig:uv_plot} shows the $uv$-coverage of our observations (top panel) and average squared visibilities for each baseline (bottom panel). Figures~\ref{Fig:data_vs_UD_1}--\ref{Fig:data_vs_UD_5} show all the obtained AMBER visibility data including the comparison to 1D CODEX and 3D CO5BOLD stellar atmosphere models. Figures~\ref{Fig:AMBER_spatfreq_plot}--\ref{Fig:CO5BOLD_spatfreq_plot} illustrate the fit of the AMBER, CODEX and CO5BOLD visibilities with a UD model for different tomographic masks.

\section{Tomographic masks}
\label{app:masks}

Here, we summarize properties (element, wavelength $\lambda_{\rm mask}$, excitation potential $\chi$, $ \log \rm gf$ and equivalent width) of spectral lines probed by the tomographic masks constructed in Sect.~\ref{Sect:tomography}.

\onecolumn
\begin{longtable}{c c c c c}
\caption{\label{tab:C1} Properties of lines contributing to the tomographic mask C1. } \\
\hline \hline
\noalign{\smallskip}
Element & $\lambda_{\rm mask}$ [$\AA$]    & $\chi$  & $\log \rm gf$ & Equivalent width [$\rm{m} \AA$ ] \\
\noalign{\smallskip}
\hline
\noalign{\smallskip}
CN  &  21343.5  &    1.08  &  -1.93  &  77.06  \\
CN  &  21435.6  &    1.05  &  -1.79  &  96.16  \\
CN  &  21441.8  &    1.30  &  -1.93  &  48.99  \\
CN  &  21858.2  &    1.13  &  -1.89  &  72.99  \\
CN  &  21905.9  &    1.25  &  -1.87  &  58.88  \\
CN  &  21936.4  &    1.02  &  -2.14  &  63.10  \\
CN  &  21963.2  &    1.04  &  -2.28  &  48.49  \\
CN  &  21972.8  &    1.16  &  -1.84  &  73.53  \\
OH  &  21976.4  &    2.41  &  -3.79  &  99.30  \\
\ion{Fe}{I}  &  21988.7  &    6.84  &   0.15  &  94.85  \\
CN  &  22136.4  &    1.26  &  -1.67  &  76.62  \\
CN  &  22151.5  &    1.42  &  -1.89  &  39.11  \\
CN  &  22236.4  &    1.11  &  -2.26  &  42.35  \\
CN  &  22243.8  &    1.25  &  -1.75  &  70.81  \\
CN  &  22248.3  &    1.46  &  -1.88  &  36.60  \\
CN  &  22302.3  &    1.33  &  -1.66  &  68.87  \\
\ion{Ca}{I}  &  22654.3  &    4.68  &  -0.99  &  168.20 \\
\ion{Si}{I}  &  22665.8  &    6.62  &  -0.68  &  107.85 \\
CN  &  22692.4  &    1.36  &  -1.66  &  63.54  \\
CN  &  22747.7  &    1.34  &  -1.97  &  40.12  \\
\ion{Fe}{I}  &  22750.7  &    3.41  &  -5.08  &  50.91  \\
\ion{Fe}{I}  &  22812.6  &    5.79  &  -0.90  &  140.60 \\
CN  &  22817.7  &    1.40  &  -1.64  &  60.47  \\
CN  &  23066.1  &    1.30  &  -1.99  &  41.01  \\
CN  &  23079.5  &    1.31  &  -2.20  &  28.10  \\
\noalign{\smallskip}
\hline
\end{longtable}

\begin{longtable}{c c c c c}
\caption{\label{tab:C2} Properties of lines contributing to the tomographic mask C2. } \\
\hline \hline
\noalign{\smallskip}
Element & $\lambda_{\rm mask}$ [$\AA$]   & $\chi$  & $\log \rm gf$ & Equivalent width [$\rm{m} \AA$ ] \\
\noalign{\smallskip}
\hline
\noalign{\smallskip}
$\rm H_2O$  &  21130.9  &    1.13  &  -4.08  &  32.19   \\
$\rm H_2O$  &  21255.9  &    1.17  &  -3.48  &  102.95  \\
$\rm H_2O$  &  21281.5  &    1.22  &  -3.64  &  60.55   \\
$\rm H_2O$  &  21418.4  &    1.15  &  -3.57  &  95.48   \\
$\rm H_2O$  &  21493.9  &    0.95  &  -4.02  &  86.34   \\
\ion{Ti}{I} 		&  21897.5  &    1.74  &  -1.45  &  567.19  \\
\ion{Ti}{I} 		&  22004.5  &    1.73  &  -1.88  &  493.12  \\
\ion{Sc}{I} 		&  22052.5  &    1.45  &  -0.76  &  494.91  \\
\ion{Sc}{I}		&  22065.3  &    1.44  &  -0.96  &  467.74  \\
\ion{Ti}{I} 		&  22211.2  &    1.73  &  -1.77  &  512.89  \\
\ion{Ti}{I} 		&  22232.9  &    1.74  &  -1.66  &  531.34  \\
\ion{Sc}{I}		&  22266.9  &    1.43  &  -1.33  &  420.76  \\
\ion{Ti}{I}		&  22274.1  &    1.75  &  -1.76  &  511.40  \\
$\rm H_2O$  &  22283.4  &    0.95  &  -4.26  &  51.43   \\
$\rm H_2O$  &  22911.2  &    0.52  &  -4.49  &  200.13  \\
CO  		&  23109.5  &    1.51  &  -4.91  &  412.65  \\
$\rm H_2O$  &  23337.9  &    0.95  &  -4.00  &  94.21   \\
CO  		&  23458.1  &    0.43  &  -5.32  &  333.88  \\
\noalign{\smallskip}
\hline
\end{longtable}

\begin{longtable}{c c c c c}
\caption{\label{tab:C3} Properties of lines contributing to the tomographic mask C3.}\\
\hline \hline
\noalign{\smallskip}
Element & $\lambda_{\rm mask}$ [$\AA$]    & $\chi$  & $\log \rm gf$ & Equivalent width [$\rm{m} \AA$ ] \\
\noalign{\smallskip}
\hline
\noalign{\smallskip}
CO  &  22930.0  &    0.56  &  -5.22  &  650.04  \\
CO  &  22935.3  &    0.49  &  -5.26  &  665.86  \\
CO  &  22937.7  &    0.47  &  -5.27  &  670.73  \\
CO  &  22940.5  &    0.45  &  -5.28  &  675.83  \\
CO  &  22943.8  &    0.43  &  -5.30  &  680.45  \\
CO  &  22947.4  &    0.41  &  -5.31  &  684.95  \\
CO  &  22951.5  &    0.39  &  -5.32  &  689.32  \\
CO  &  22956.1  &    0.37  &  -5.33  &  693.57  \\
CO  &  22960.9  &    0.35  &  -5.35  &  697.49  \\
CO  &  22966.3  &    0.33  &  -5.36  &  701.45  \\
CO  &  22971.9  &    0.32  &  -5.37  &  705.45  \\
CO  &  22978.1  &    0.30  &  -5.39  &  708.94  \\
CO  &  22984.9  &    0.28  &  -5.40  &  712.11  \\
CO  &  22991.7  &    0.27  &  -5.42  &  715.49  \\
CO  &  22999.0  &    0.25  &  -5.44  &  718.18  \\
CO  &  23006.8  &    0.24  &  -5.45  &  720.90  \\
CO  &  23015.0  &    0.22  &  -5.47  &  723.45  \\
CO  &  23023.5  &    0.21  &  -5.49  &  725.68  \\
CO  &  23032.6  &    0.19  &  -5.50  &  727.92  \\
CO  &  23041.9  &    0.18  &  -5.52  &  729.47  \\
CO  &  23051.8  &    0.17  &  -5.54  &  731.22  \\
CO  &  23061.8  &    0.16  &  -5.56  &  732.44  \\
CO  &  23072.2  &    0.14  &  -5.58  &  733.50  \\
CO  &  23083.1  &    0.13  &  -5.60  &  734.39  \\
CO  &  23094.4  &    0.12  &  -5.62  &  734.94  \\
CO  &  23106.1  &    0.11  &  -5.64  &  734.95  \\
CO  &  23118.3  &    0.10  &  -5.67  &  734.79  \\
CO  &  23130.8  &    0.09  &  -5.69  &  734.46  \\
CO  &  23143.8  &    0.08  &  -5.72  &  733.40  \\
CO  &  23157.1  &    0.07  &  -5.74  &  731.97  \\
CO  &  23170.8  &    0.06  &  -5.77  &  730.37  \\
CO  &  23185.0  &    0.06  &  -5.80  &  728.24  \\
CO  &  23199.5  &    0.05  &  -5.83  &  725.72  \\
CO  &  23214.6  &    0.04  &  -5.86  &  722.66  \\
CO  &  23221.3  &    0.82  &  -4.74  &  648.38  \\
CO  &  23230.4  &    0.04  &  -5.90  &  718.86  \\
CO  &  23233.7  &    0.69  &  -4.81  &  677.66  \\
CO  &  23237.3  &    0.67  &  -4.83  &  682.15  \\
CO  &  23241.4  &    0.65  &  -4.84  &  686.34  \\
CO  &  23245.7  &    0.03  &  -5.93  &  714.70  \\
CO  &  23250.8  &    0.61  &  -4.87  &  694.11  \\
CO  &  23256.4  &    0.60  &  -4.88  &  697.87  \\
CO  &  23261.7  &    0.03  &  -5.98  &  709.44  \\
CO  &  23267.8  &    0.56  &  -4.91  &  704.59  \\
CO  &  23274.3  &    0.55  &  -4.92  &  707.57  \\
CO  &  23281.2  &    0.53  &  -4.94  &  710.39  \\
CO  &  23288.4  &    0.51  &  -4.96  &  713.23  \\
CO  &  23295.9  &    0.50  &  -4.97  &  715.75  \\
CO  &  23304.3  &    0.49  &  -4.99  &  717.77  \\
CO  &  23312.9  &    0.47  &  -5.01  &  719.98  \\
CO  &  23321.9  &    0.46  &  -5.02  &  721.87  \\
CO  &  23331.1  &    0.44  &  -5.04  &  723.43  \\
CO  &  23341.3  &    0.43  &  -5.06  &  724.84  \\
CO  &  23351.4  &    0.42  &  -5.08  &  725.91  \\
CO  &  23362.2  &    0.41  &  -5.10  &  726.83  \\
CO  &  23367.9  &    0.00  &  -6.33  &  655.14  \\
CO  &  23373.4  &    0.40  &  -5.12  &  727.42  \\
CO  &  23406.2  &    0.00  &  -6.56  &  617.62  \\
CO  &  23421.2  &    0.35  &  -5.21  &  727.15  \\
CO  &  23434.3  &    0.35  &  -5.23  &  726.23  \\
CO  &  23447.6  &    0.34  &  -5.26  &  724.62  \\
\noalign{\smallskip}
\hline
\end{longtable}

\end{appendix}

\end{document}